\documentclass{aa}

\DeclareRobustCommand{\VAN}[3]{#2}
\let\VANthebibliography\thebibliography
\def\thebibliography{\DeclareRobustCommand{\VAN}[3]{##3}\VANthebibliography}

\usepackage[colorlinks=true,
    linkcolor=blue, citecolor=blue, filecolor=blue, urlcolor=blue]{hyperref}
\usepackage{xurl}
\usepackage{caption}
\usepackage{subcaption}
\usepackage{mhchem}
\usepackage{graphicx}	
\usepackage{amsmath}	
\usepackage{amssymb}	
\usepackage{array}
\usepackage{multirow}
\usepackage{bigdelim}
\usepackage[varg]{txfonts}
\usepackage{xcolor}
\usepackage{lscape}
\usepackage{booktabs}

\newcommand{\dd}{\mathrm{d}}

\begin{document}


\title{The Astrochemistry Low-energy Electron Cross-Section (ALeCS) database I.}
\subtitle{Semi-empirical electron-impact ionization cross-section calculations and ionization rates}
\titlerunning{ALeCS database}
\authorrunning{Gaches et al.}

\author{Brandt A. L. Gaches
        \inst{1}\fnmsep\inst{2}\thanks{E-mail: brandt.gaches@chalmers.se} 
        \and
        Tommaso Grassi\inst{3}
        \and
        Stefan  Vogt-Geisse\inst{4}
        \and
        Giulia M. Bovolenta\inst{4,5}
        \and
        Claire Vallance\inst{6}
        \and
        David Heathcote\inst{6}
        \and
        Marco Padovani\inst{7} 
        \and
        Stefano Bovino\inst{8,9,10}
        \and
        Prasanta Gorai\inst{1,11}
        }
\institute{
Department of Space, Earth and Environment, Chalmers University of Technology, Gothenburg SE-412 96, Sweden
\and
Center of Planetary Systems Habitability, The University of Texas at Austin,  USA
\and
Max Planck Institute for Extraterrestrial Physics, Giessenbachstrasse 1, 85748 Garching bei M\"{u}nchen, Germany
\and
Departamento de Físico-Química, Facultad de Ciencias Químicas, Universidad de Concepción, Concepción, Chile
\and
Atomistic Simulations, Italian Institute of Technology, 16152 Genova, Italy
\and
Department of Chemistry, University of Oxford, Chemistry Research Laboratory, 12 Mansfield Road, Oxford OX1 3TA, U.K
\and
INAF–Osservatorio Astrofisico di Arcetri, Largo E. Fermi 5, 50125 Firenze, Italy
\and
Departamento de Astronom\'{i}a, Facultad Ciencias F\'{i}sicas y Matem\'{a}ticas, Universidad de Concepci\'{o}n Av. Esteban Iturra s/n Barrio Universitario, Casilla 160, Concepción, Chile
\and
INAF—Istituto di Radioastronomia, Via Gobetti 101, I-40129 Bologna, Italy
\and
Chemistry Department, Sapienza University of Rome, P.le A. Moro, 00185 Rome, Italy
\and
Department of Chemistry and Molecular Biology, University of Gothenburg, 41296 Gothenburg, Sweden
}

\date{Accepted XXX. Received YYY; in original form ZZZ}

\abstract
{Electron-molecule interaction is a fundamental process in radiation-driven chemistry in space, from the interstellar medium to comets. Therefore, knowledge of interaction cross sections is key. There have been a plethora of both theoretical and experimental studies of total ionization cross sections spanning from diatomics to complex organics. However, data is often spread over many sources, or not public or readily available.}
{We introduce the Astrochemistry Low-energy Electron cross-section (ALeCS) database, a public database for electron interaction cross sections and ionization rates for molecules of astrochemical interest. In particular, in this work, we present the first data release comprising total ionization cross sections and ionization rates for over 200 neutral molecules.}
{We include optimized geometries and molecular orbital energies at various levels of quantum chemistry theory. Further, for a subset of the molecules we have calculated ionization potentials. We compute total ionization cross sections using the binary-encounter Bethe model and screening-corrected additivity rule, and ionization rates and reaction network coefficients for molecular cloud environments.}
{We present the cross sections and reaction rates for $>$200 neutral molecules ranging from diatomics to complex organics, with the largest being \ce{C14H10}. We find that the screening-corrected additivity rule cross sections generally significantly overestimate experimental total ionization cross sections. We demonstrate that our binary-encounter Bethe cross sections agree well with experimental data. We show that the ionization rates scale roughly linearly with the number of constituent atoms in the molecule.}
{We introduce and describe the public ALeCS database. For the initial release, we include total ionization cross sections for $>$200 neutral molecules and several cations and anions calculated with different levels of quantum chemistry theory, the chemical reaction rates for the ionization, and network files in the formats of the two most popular astrochemical networks, the Kinetic Database for Astrochemistry and UMIST. The database will be continuously updated for more molecules and interactions.}

\keywords{Astrochemistry - Molecular data - Molecular processes -  ISM: molecules -  ISM: cosmic rays}

\maketitle

\section{Introduction}
Observational studies of molecular gas within the Milky Way have revealed a diverse zoo of about 300 molecules, from simple diatomics such as \ce{H2} and \ce{CO}, to ever more complex molecules, such as \ce{NC4NH+} \citep{Agundez2023}, \ce{NH2C(O)CH2OH} \citep[{\em syn}-glycolamide,][]{Rivilla2023}, \ce{2-C9H7CN} \citep[2-Cyanoindene,][]{Sita2022}, and \ce{H2CCCHC3N} \citep[cyanoacetyleneallene,][]{Shingledecker2021}. Due to the cold temperatures of the interstellar medium ($\sim$10-50 K), much of the gas-phase chemistry is driven through ion-neutral initiated chemistry \citep{larsson2012, tielens2013, vandishoeck2014}. In regions shielded from external radiation, highly energetic charged particles, known as cosmic rays, in particular protons and secondary electrons, and photons, provide the primary ionization source \citep{umebayashi1981}. Finally, chemical models of ices irradiated by energetic particles or X-ray radiation, such as interstellar dust grains and comets, necessarily need to include electron-impact ionization cross sections for as many molecules as possible in order to account for electron production and subsequent interactions \citep[e.g.][]{Shingledecker2020}.

Astrochemical models describe the evolution of vast chemical networks over time for a wide range of physical environments, including the effects of gas density, temperature, atomic abundances, and radiation environments. There are two key astrochemical databases containing the required reaction rate  coefficients, {\sc KIDA}\footnote{\url{https://kida.astrochem-tools.org/}} \citep{Wakelam2012} and {\sc UMIST}\footnote{\url{http://udfa.ajmarkwick.net/}} \citep{McElroy2013}. However, for cosmic-ray ionization, there is a substantial paucity in reaction rate data, with {\sc KIDA} and {\sc UMIST} combined only reporting reaction rate coefficients for H, He, C, O, N, \ce{H2}, \ce{N2}, and \ce{CO}. In astrochemical models, these reaction rates are expressed in the form 
\begin{equation}
    \zeta_{m, T} = c_{m,T} \zeta_{\ce{H2}, T},
\end{equation}
where $c_{m}$ is a scaling factor that relates the rate of the reaction of interest to the total ionization rate, $\zeta_{\ce{H2}, T}$, of \ce{H2}. The coefficients are computed with respect to some reference total \ce{H2} ionization rate, typically $\zeta_{\ce{H2},0} \approx 3\times10^{-17}$ s$^{-1}$, often referred to as the ``canonical'' or ``fiducial'' ionization rate. 

The coefficients  $c_{m,T}$ in KIDA and UMIST date back to early studies from the 1970s and 1980s \citep{Cravens1975, Cravens1978, Glassgold1974, Black1975}. For molecules containing atoms beyond hydrogen and helium, computed ionization rates often use scaling relationships between the high-energy cross sections of the molecules and assumed a Voyager-like proton cosmic-ray spectrum. However, there has been substantial development in the theoretical calculation and experimental measurements of these cross sections. Further, it has been demonstrated that the spectrum of secondary non-thermal electrons is highly sensitive to the local proton cosmic-ray spectral shape \citep{Ivlev2021}. Observations of the ionization rate have demonstrated that the \ce{H2} ionization rate within molecular clouds is more commonly around $10^{-16}$\,s$^{-1}$ \citep{caselli1998, vandertak2000, Neufeld2010, indriolo2012, indriolo2015, neufeld2017, Sabatini2020, Luo2023a, Luo2023b, Sabatini2023}, with trend of decreasing towards increasing column density, consistent with a combination of energy losses and possible diffusive transport \citep[e.g.][]{padovani2009, Padovani2018, silsbee2019, Phan2023}. 

\urldef{\NISTURL}\url{https://www.nist.gov/pml/electron-impact-cross sections-ionization-and-excitation-database}
There have now been a sizable number of studies reporting computed values for the total electron-impact ionization cross sections of a variety of molecules. Public databases such as PlasmaData\footnote{\url{http://plasma.mathboylinlin.com/}} \citep{Zhong2021},  BEAMDB\footnote{\url{http://servo.aob.rs/emol/}} \citep{Marinkovic2017}, the US-based National Institute Standards and Technology (NIST) Electron-Impact cross sections for Ionization and Excitation Database\footnote{\NISTURL}, and the Japan-based National Institute for Fusion Science (NIFS) datatabase\footnote{\url{https://dbshino.nifs.ac.jp/nifsdb/}} 
have compiled the cross sections for a wide range of molecules, often for use in plasma physics, medicine, or other industry applications. The latter of these databases comprises a significant number of experimental and evaluated ionization cross sections. Recently \citet{Heathcote2018} and \citet{Zhou2019} presented large datasets of calculated and experimental cross sections, respectively, for a wide range of molecules of astrochemical interest. These datasets and databases represent a significant advancement in the availability of cross-section data for astrochemical use. However, they generally do not include molecules containing heavier atoms, and also do not report on computed and recommended reaction rate coefficients. Furthermore, databases comprising a large number of ionization cross sections often either do not use a standard evaluation process or do not provide the data in a readily accessible way.

There have been a number of different electron-impact total ionization cross sections introduced in the literature \citep[see e.g.][]{Kim1997, Deutsch2000, Blanco2010}. For large molecules, in which the computing electronic structure can become untenable, various different additivity rules have been introduced which introduce a sort of algebra for adding atomic ionization cross sections, atomic or molecular orbits. One of the most important models is the binary-encounter dipole and its simplification, the binary-encounter Bethe cross section \citep{Kim1994, Kim1997}, detailed below in Section \ref{sec:BEB}. It has performed well compared to experimental data \citep[e.g.]{Kim1997, Zhou2019, Zhong2021} has thus become a standard. All of the above databases also present the binary-encounter Bethe cross sections as a major result.

In this paper, we present a new public database containing electron ionization cross sections and cosmic-ray ionization rate coefficients for over 200 molecules of astrochemical interest. We report the cross sections for low-energy electrons (10~eV $\leq E_e \leq $ 5000~eV) for each of these molecules using both a screening-corrected additivity rule and the binary-encounter Bethe model cross section, the molecular orbital binding and kinetic energies utilized, and recommended total reaction rate coefficients for molecular cloud environments.  Crucially, the cross sections and coefficients are evaluated for nearly all molecules using the same procedure, providing homogeneous datasets to enable better comparisons and consistency.

\section{Methods}
We present here the methods for the different approaches of calculating the total electron-impact ionization cross sections, then quantum chemistry computations performed, and the calculations of the resulting ionization rates. In brief, we calculate three different models of the electron-impact total ionization rate: the screening-corrected additivity rule (SCAR), binary-encounter Bethe (BEB), and damped-binary-encounter Bethe (dBEB) models. In particular, we explore the accuracy of the SCAR method, since it may become more applicable for larger molecules due to the computational expense required for accurately computing BEB-related cross sections. The latter models require knowledge of the electronic structure, while the former only requires knowledge of the geometry and a basis set of electron-atom ionization cross sections. For our molecule selection, we choose the primary molecules used in astrochemical networks with reaction rate data on the kinetics database KIDA. 

We emphasize that the total ionization cross sections below are for single ionization events. At high energies, multiple ionization events can occur, in particular through Auger ionization. \citet{nishimura1999} estimated the possible impact of multiple ionizations by doubling the contribution of the cross section from inner shell electrons and found their inclusion shifted the peak by 5\% and towards higher impact energies. Multiple ionization event contributions are beyond the scope of this initial release, and will be considered for future releases. Further, molecules are assumed to be ionized from their ground state.

\subsection{Electron-atom ionization cross sections}
\begin{figure*}[htb!]
    \centering
    \includegraphics[width=\textwidth]{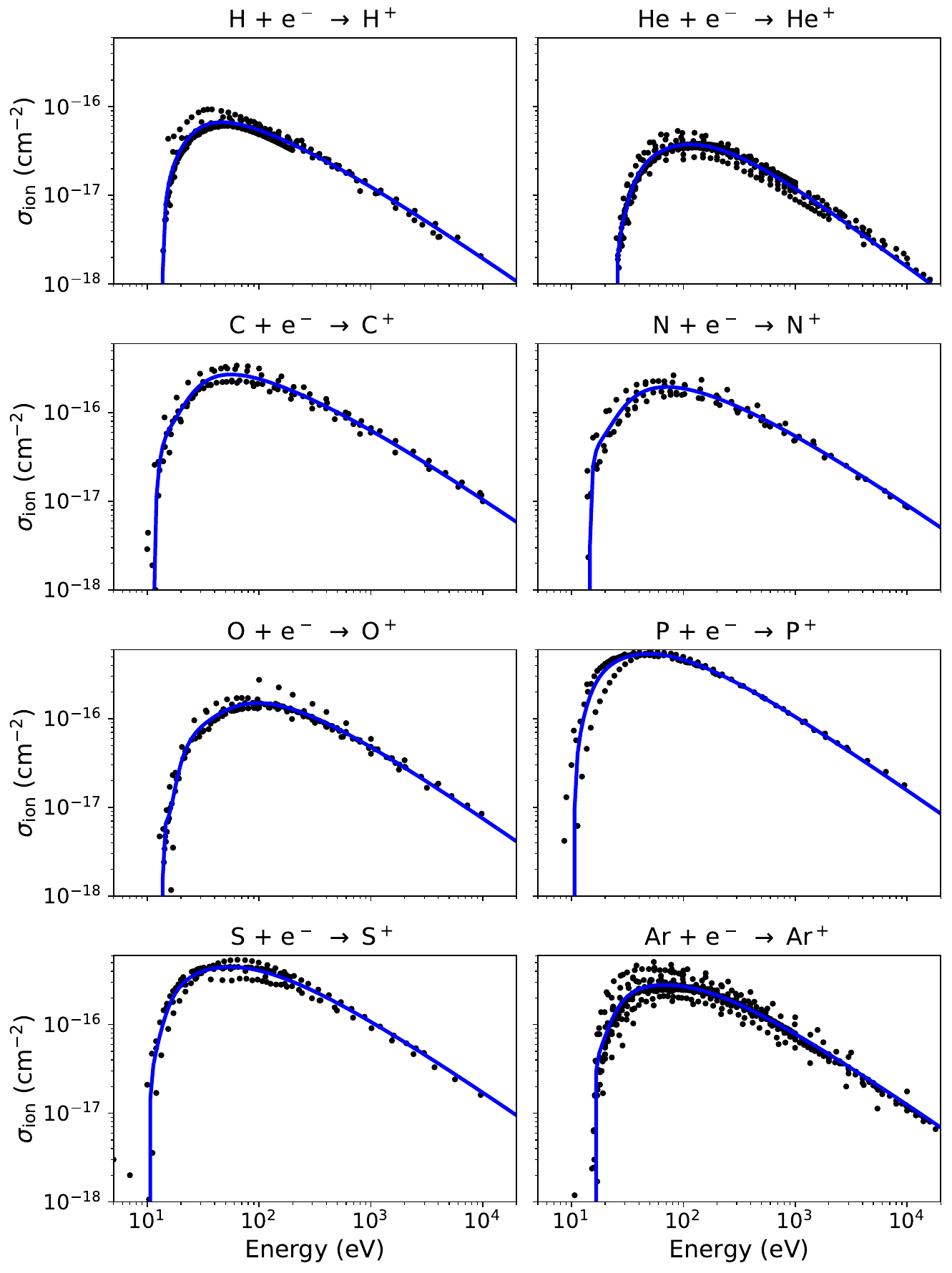}
    \caption{\label{fig:atomicData}Electron ionization cross sections. Solid lines give the fits from Equation \ref{eq:atomFit} and the black points are the data from the NIFS database.}
\end{figure*}

The semi-empirical method we use requires a basis of electron-atom ionization cross sections. We obtain these cross sections by fitting a polynomial to experimental data, from the NIFS AMIDIS-ION\footnotemark[5] database, where the electron-atom ionization cross section takes the form:
\begin{equation}\label{eq:atomFit}
\sigma_{i}(E) = 
\begin{cases}
0 & \text{for} \,\, E_e < {\rm IP}_i \\
a_0^2 \left ( \frac{x-1}{x}\right ) \left [ c_0^2 \left ( \frac{\ln x}{x} \right ) + \sum_{k=1}^5 \frac{c_k}{x^k} \right ] & \text{for} \,\, E_e \ge {\rm IP}_i,
\end{cases}
\end{equation}
where $a_0$ is the Bohr radius, $E_e$ is the electron energy, IP$_i$ is the ionization potential of the atomic species $i$, $x = E_e/{\rm IP}_i$, and $c_k$ are fitting coefficients. Figure \ref{fig:atomicData} shows the experimental data used and the fitted cross sections. Table \ref{tab:atmcoeffs} shows the results of the fits of Equation \ref{eq:atomFit} for each atom.

\begin{table*}
    \centering
    \caption{\label{tab:atmcoeffs} Electron-atom ionization cross section fitted coefficients for Equation \ref{eq:atomFit}. Coefficients are given in the form $a(b)$ = $a \times 10^b$. }
    \begin{tabular}{l|cccccc}
    \toprule\toprule
    Atom & $c_0$ & $c_1$ & $c_2$ & $c_3$ & $c_4$ & $c_5$ \\
    \midrule
    H & 2.813905(0) & $-$1.101730(0) & 1.838647(1) & $-$3.577323(1) & 2.554819(1) & $-$2.358867(0) \\
    He & 1.313597(0) & 1.261778(1) & $-$5.359803(1) & 1.153092(2) & $-$1.239072(2) & 5.145647(1) \\
    C & 7.575591(0) & $-$5.795490(1) & 2.264768(2) & $-$6.500235(2) & 9.067897(2) & $-$4.246887(2) \\
    N & 6.180613(0) & $-$2.659850(1) & 6.022187(1) & $-$1.036028(2) & 1.070141(2) & $-$1.959618(1) \\
    O & 5.400010(0) & 4.102912(0) & $-$1.951275(2) & 6.362247(2) & $-$7.754242(2) & 3.372681(2) \\
    P & 8.463675(0) & 3.806864(1) & $-$3.042532(2) & 8.754706(2) & $-$1.029666(3) & 4.429356(2) \\
    S & 9.674459(0) & $-$5.296533(1) & $-$8.623544(1) & 6.587089(2) & $-$9.794655(2) & 4.823979(2) \\
    Ar & 6.618024(0) & 9.689090($-$1) & $-$1.248942(2) & 5.386474(2) & $-$7.977861(2) & 4.114582(2) \\
    \bottomrule
    \end{tabular}
\end{table*}

\subsection{Semi-empirical cross-section calculation}
\subsubsection{Additivity Rule}
The simplest semi-empirical cross section is the simple additivity rule (AR). Here, the total molecule ionization cross section of species, $m$, is
\begin{equation}
    \sigma_{m, {\rm AR}}(E) = \sum_i \sigma_i(E),
\end{equation}
where the summation is over the different atoms, $i$. Note that this ignores all possible overlap of the atomic cross sections and electronic structure and will always provide an overestimate of the ionization cross section.

\subsubsection{SCAR Method}
The cross sections are computed using the \emph{screening-corrected additivity rule} (SCAR) \citep{Blanco2003, Blanco2010}. This method enables the quick computation of molecular cross sections using just a pre-computed basis of electron-atom ionization cross sections. SCAR is an additivity rule which takes into account the effective geometrical overlap of the atomic electron cross sections. The cross section for a specific molecule, $m$, is given by
\begin{equation}
    \sigma_{m, {\rm SCAR}}(E) = \sum_i s_i \sigma_i(E),
\end{equation}
where the sum is carried out over all constituent atoms and $\sigma_i(E)$ is the electron-atom ionization cross section for atom, $i$. Under the simplest additivity rules, as shown above, $s_i = 1$. For the SCAR method, the additivity coefficients, $s_i$, are
\begin{equation}\label{eq:SCARCoeff}
s_i = \sum_{k=1}^{N_k} \frac{\left ( -1\right )^{k+1} \varepsilon_i^{(k)}}{k!},
\end{equation}
where $N_k$ is the number of perturbation terms included. The coefficients, $\varepsilon_i^{(k)}$, follow the recursive relation
\begin{align}\label{eq:recursive}
\varepsilon_i^{(1)} &= 1 \\
\varepsilon_i^{(k)} &= \frac{N_k - k + 1}{N_k - 1} \sum_{\rm j(\neq i)} \frac{\sigma_j \varepsilon_j^{(k-1)}}{\alpha_{ji}}\,, k = 2,\ldots,N\,,
\end{align}
where $\alpha_{ji} = \max(4\pi r_{ij}^2, \sigma_i, \sigma_j)$, and $r_{ij}$ is the distance between atoms $i$ and $j$. The terms in Equation \ref{eq:SCARCoeff} amount to higher-order correction factors accounting for the overlap of the cross sections of all the individual items. Due to the recursive nature of the coefficients (Eq. \ref{eq:recursive}), including higher-order correction terms leads to an exponential increase in computational cost. However, this is alleviated by building a dictionary cache during the recursion to avoid recomputing the same screening terms. This leads to a sub-linear increase in computing time against maximum $k$-atom screening correction included.

\subsubsection{BEB cross sections}\label{sec:BEB}
The semi-empirical binary-encounter Bethe (BEB) cross section was developed as a simplification of the binary-encounter dipole cross section \citep{Kim1994,Hwang1996}. This cross section has been found to provide a reasonable match with experimental data, and requires only knowledge of the molecular orbital energies. BEB cross section are the base of numerous cross-section databases, in particular the NIST database. The BEB cross section is defined as:
\begin{align}
    \sigma_{m, {\rm BEB}} = \sum_{\ell}  & \left [ \frac{S_\ell}{t_\ell + (u_\ell + 1)}\right ] \Biggl [ \frac{\ln t_\ell}{2} \left ( 1 - \frac{1}{t_\ell^2} \right ) + \\ & \left ( 1 - \frac{1}{t_\ell} - \frac{\ln t_\ell}{t_\ell + 1}\right ) \Biggr ],
\end{align}
where the sum is over orbitals, indexed $\ell$, $t_\ell =  E_e/B_\ell$, $u_\ell = U_\ell/B_\ell$ and $S_\ell = 4\pi a_0^2 n_\ell \left ( \frac{R_{\infty}}{B_\ell}\right )^2$. Here, $B_\ell$ and $U_\ell$ are the orbital binding and orbital kinetic energies of the ejected electron, respectively, $n_\ell$ is the orbital occupation number, and $R_{\infty}$ is the Rydberg constant. The parameters $B_\ell$ and $U_\ell$ are generally computed using electronic structure methods. 

We also propose a modified version of the BEB cross section (dBEB) to dampen the impact of orbitals with binding energies greater than the ionization potential. In dBEB cross-section formulation, the $B_\ell$ is scaled by an exponential
function such that the modified binding energy of orbital $\ell$ is given by:
\begin{equation}
    B_\ell^\prime = B_\ell e^{-(1 - B_\ell/{\rm IP})},
\end{equation}
As we demonstrate below, this prevents the BEB cross section from over-estimating with respect to experimental values, which are generally upper limits. In general, BEB cross sections underpredict experimental results at higher energies since they only account for single ionizations, while experimental data, where the measured signal is generally the ion current, includes contributions from multiple ionization events which contribute more to the measured signal per event than single ionizations.

\subsection{Quantum chemistry computations}
We selected a set of 156 neutral species including atoms and molecules, for the calculation of BEB and dBEB cross sections. The initial structures are taken from NIST Computational Chemistry Comparison and Benchmark DataBase\footnote{\url{https://cccbdb.nist.gov/}} (CCCBDB) database \citep{NIST_CCCBDB} or ChemSpider\footnote{\url{https://www.chemspider.com/}}. The equilibrium geometries were obtained at MP2/aug-cc-pVTZ method and basis set. To validate the accuracy of this level of theory we also optimized a group of 50 structures using the highly accurate DF-CCSD(T)-F12/cc-pVDZ-F12 (U-CCSD(T)-F12/cc-pVDZ-F12 for radical species). The set of electron binding energy (eBE), $B_\ell$, taken as the negative of the energy of orbital $\ell$, is computed by means of electron propagator theory (EPT) in its P3+ implementation along with aug-cc-pVTZ basis. In EPT,  electron correlation is taken into account, thus providing a more accurate value for $B_\ell$, compared to Hartree-Fock (HF) canonical orbital energies. For the computation of the BEB cross sections, we use the EPT-corrected orbital energies if the related pole strength is greater than 0.8, or else we use the HF canonical orbital energies as an estimate of the $B_\ell$. Since the lowest $B_\ell$ -- namely the IP, according to Koopmans' theorem --  is the predominant contribution to the BEB cross section, the accuracy of the eBE plays a significant role.

We computed vertical IPs for 148 molecules in our sample, using the Density Functional Theory (DFT) method CAM-B3LYP/aug-cc-pVQZ, employing the D3 empirical dispersion. Furthermore, to validate the accuracy of the IPs we compare the DFT results to highly accurate values that we computed at  CCSD(T) paired with complete basis set (CBS) extrapolation using cc-pVXZ (X=D,T) basis functions, for a subset of 56 species. This subset consists of molecules with 2-4 atoms. 

Computations were performed using software packages  {\sc Gaussian16}\footnote{\url{http://gaussian.com/}} \citep{g16}, {\sc Molpro}\footnote{\url{https://www.molpro.net/}} \citep{MOLPRO-WIREs, Werner2020}, and  {\sc Psi4}\footnote{\url{https://psicode.org/}} (\citealt{Psi4}, interfaced with the {\sc QCFractal}\footnote{\url{https://github.com/MolSSI/QCFractal}} infrastructure, see \citealt{smith_molssi_2020}). Details of the different electronic structure methods are presented in Table \ref{tab:methods}.

\begin{table*}
    \centering
    \caption{\label{tab:methods} Summary of quantum chemistry methods and citations.}
   \begin{tabular}{r|p{0.4\textwidth}|p{0.4\textwidth}}
    \toprule\toprule
    \multirow{2}{*}{Basis sets} & aug-cc-pVXZ (X=T,Q) & \citet{Dunning1989}, \citet{DunningT.H.2001} \\
    & Complete Basis Set (CBS) extrapolation & \citet{helgaker_basis-set_1997}\\ 
  \hline
    \multirow{4}{*}{Gaussian16} & M\o{}ller-Plesset second order (MP2) & \citet{MP2}, \citet{FRISCH1990275}, \citet{FRISCH1990281} \\
      & CAM-B3LYP & \citet{Becke1993}, \citet{YANAI200451} \\
      & Grimme's dispersion (D3) & \citet{Grimme2010} \\
      & Electron-propagator Theory (EPT), P3+ & \citet{Migdal1968}, \citet{Ortiz2005} \\
  \hline
   \textsc{Psi4} & Coupled Cluster (CCSD(T)) & \citet{raghavachari_fifth-order_1989}\\
   \hline
    \multirow{1}{*}{\textsc{Molpro}} & Explicitly correlated Density Fitted CCSD(T) (DF-CCSD(T)-F12) & \citet{gyorffy_analytical_2018} \\
    \bottomrule
    \end{tabular}
\end{table*}

Finally, we include molecular geometries and orbitals from \citet{Heathcote2018}, which comprises 141 molecules including 6 anions and 18 cations. These were computed at the HF/aug-cc-pVTZ level, with a small subset at the MP2/aug-cc-pVTZ level (\ce{C4H}, \ce{C5H}, \ce{C5N}, and \ce{C6H}, see \citealt{Heathcote2018} for details). \ce{H2} is computed at the CCSD(T)/aug-cc-pVTZ level. We also include the non-published calculated data for cations and anions which were previously excluded since there is a lack of experimental data for these molecules to benchmark against, and so the method of calculation could not be confirmed. As such, we exclude the anions and cations from the reaction rate plots, although they are included in the full database.

\subsection{Low-energy electron spectrum and ionization rate}\label{sec:crion}
Low-energy electrons in molecular clouds are primarily generated as secondary particles produced by ionization of \ce{H2} by primary cosmic-ray protons or electrons. We obtain depth-dependent electron spectra, e.g., $j_{e}(E_e, N_{\ce{H2}})$ where $N_{\ce{H2}}$ is the \ce{H2} column density in the cloud following the new rigorous prescription in \citet{Ivlev2021}, with corrections for \ce{H2}-excitation and new \ce{H2}-electron cross sections \citep{Padovani2022}.

The depth-dependent electron ionization rate of a molecule, $m$, is given by adding the contributions of primary protons, $\zeta_{p, m}(N_{\ce{H2}})$, using the approximation ${m_e}E_p \approx {m_p}E_e$,  
\begin{equation}
    \zeta_{p,m}(N_{\ce{H2}}) = 2\pi \int j_{p}(E_p, N_{\ce{H2}}) \,\sigma_{m}\left[(m_e/m_p) E_p\right]\,\dd E_p,
\end{equation}
primary electrons,
\begin{equation}
    \zeta_{e,m}(N_{\ce{H2}}) = 2\pi \int j_{e}(E_e, N_{\ce{H2}}) \,\sigma_{m}(E_e)\,\dd E_e,
\end{equation}
and secondary electrons from both primary protons and electrons
\begin{equation}
    \zeta_{se,m}(N_{\ce{H2}}) = 4\pi \int j_{se}(E_e, N_{\ce{H2}}) \,\sigma_{m}(E_e)\,\dd E_e,
\end{equation}
where $se$ represents secondary electrons. The factors of $2\pi$ and $4\pi$ account for fluxes which are assumed to be plane parallel and isotropic, respectively. The total ionization rate for a molecule is then, $\zeta_{m} = \zeta_{p,m} + \zeta_{e,m} + \zeta_{se,m}$. The factor of $4\pi$ comes from treating the secondary electrons as an isotropic local source. 

We use the ``High'' primary proton and proton-induced secondary electron spectra as a function of column density from \citet{Padovani2022}, who computed the electron spectrum down to 1 eV following the recent more rigorous theory of \citet{Ivlev2021}. The High proton spectrum has been calibrated to match diffuse gas observations of the \ce{H2} ionization rate, which are not reproduced with a Voyager-like spectrum \citep{Ivlev2015, Padovani2018}. We also use an interstellar primary electron spectrum from \citet{Padovani2018}, and their induced secondary electrons. The secondary electron computation assumes that the gas is fully molecular and includes energy losses from Coloumb interactions, as well as for \ce{H2} ionizations and electronic and rovibrational excitations. We define our column-dependent total ionization rate coefficient for species $m$, $c_{m, T}(N_{\ce{H2}})$ from 
\begin{equation}\label{eq:coeffs}
    \zeta_{m, T}(N_{\ce{H2}}) = c_{m, T}(N_{\ce{H2}})\,\zeta_{\ce{H2},T}(N_{\ce{H2}})\,,
\end{equation}
where $\zeta_{\ce{H2},T}(N_{\ce{H2}})$ is the total \ce{H2} ionization rate including all primary, secondary, and tertiary processes. We report the column-density average coefficient, $\bar{c}_{m,T}$, for cloud column density range, $N_{\ce{H2}} \approx 10^{20} - 10^{23}$~cm$^{-2}$, although we note that the coefficients only marginally scale with column density. Hereafter, we denote $\bar{c}_{m,T}$ as $c_{m,T}$, due to the marginal scaling with column density. In general, since the BEB cross sections perform better than the SCAR data in comparisons against experimental data, we report here only the coefficients using the BEB cross section, although all cross sections are available in the database.

\section{Results}
\begin{figure}
    \centering
    \includegraphics[width=0.5\textwidth]{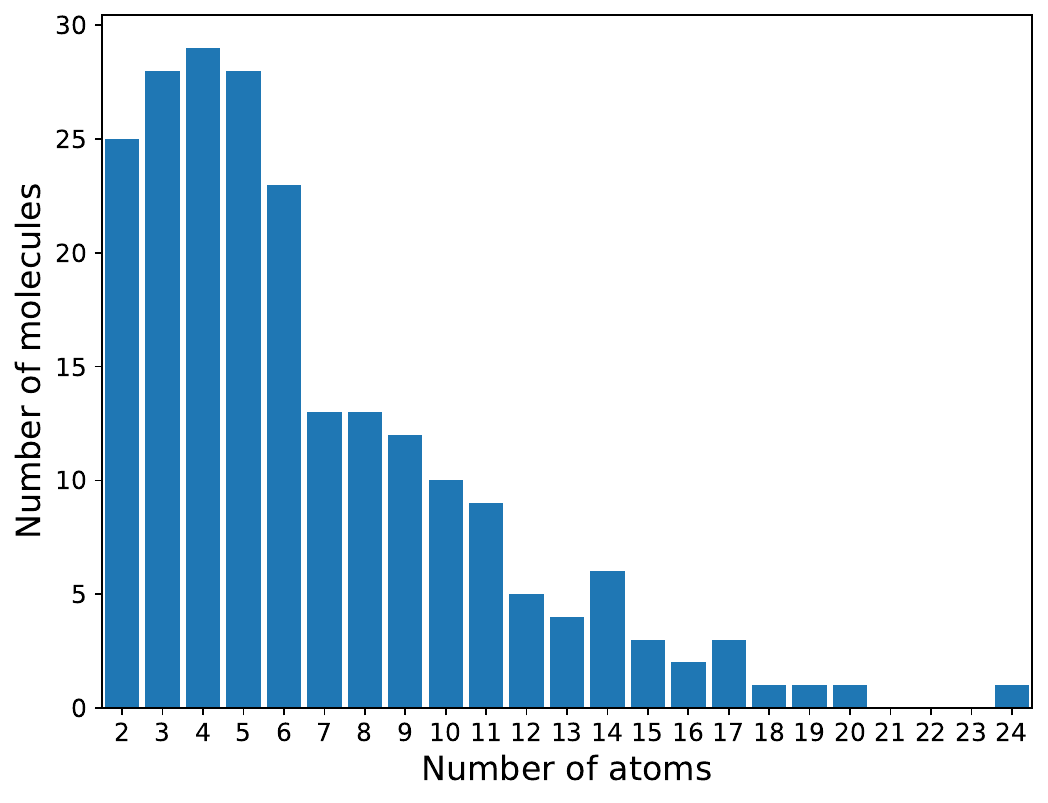}
    \caption{\label{fig:molStats}Distribution of the number of atoms in each molecule for this data release.}
\end{figure}

In total, we have computed the structure, orbitals, and cross sections for 202 unique (by composition, not counting isomers) neutral molecules ranging in size from two to 24 atoms including data augmented by the results from \citet{Heathcote2018}. Figure \ref{fig:molStats} shows the distribution of the number of atoms for the molecules in our database; while most of the molecules have fewer than six atoms, we include some with up to 24 atoms (\ce{C14H10}). In cases where KIDA does not specify an isomer, but multiple isomers exist, we take the  most stable isomer. We detail below a summary of the results, with all of the data available online in the public database, \url{https://github.com/AstroBrandt/ALeCS}.
\begin{sidewaystable*}[b]
    \centering
    \caption{\label{tab:molecules}Unique molecules included (202 total) in this database, sorted by their number of constituent atoms.}
    \begin{tabular}{|*{11}{c|}}
    \hline
    Atoms & 2 atoms & 3 atoms & 4 atoms & 5 atoms & 6 atoms & 7 atoms & 8 atoms & 9 atoms & 10-15 atoms & 15+ atoms \\
    \hline\hline
    \ce{H} & \ce{C2} & \ce{C2H} & \ce{C2H2} & \ce{C2H2O} &
  \ce{C2H4} & \ce{C2H4O} & \ce{C2H3NH2} & \ce{C2H5CN} &
  \ce{C10H2} & \ce{C14H10} \\
 \ce{He} & \ce{CF+} & \ce{C3} & \ce{C2N2} & \ce{C2H3} &
  \ce{C3H3} & \ce{C2H5} & \ce{C2H6} & \ce{C2H5OH} &
  \ce{C2H5CHO} & \ce{C5H10} \\
 \ce{Ar} & \ce{CH} & \ce{CCN} & \ce{C3H} & \ce{C2HNO} &
  \ce{C5H} & \ce{C6H} & \ce{C3H5} & \ce{C3H6} &
  \ce{C2H5OCH3} & \ce{C5H11} \\
 \ce{C} & \ce{CH+} & \ce{CCO} & \ce{C3H+} & \ce{C3H2} &
  \ce{C5N} & \ce{CH2CCH2} & \ce{C4H4} & \ce{C4H5} &
  \ce{C3H5CN} & \ce{C5H12} \\
 \ce{N} & \ce{CN} & \ce{CH2} & \ce{C3N} & \ce{C4H} &
  \ce{CH2CCO} & \ce{CH2CHCN} & \ce{C7H} & \ce{C5H4} &
  \ce{C3H7} & \ce{C6H10} \\
 \ce{O} & \ce{CO} & \ce{CNO} & \ce{C3O} & \ce{C5} &
  \ce{CH2NH2} & \ce{CH3C2H} & \ce{CH2CHCHO} & \ce{C8H} &
  \ce{C3H7CN} & \ce{C6H11} \\
 \ce{S} & \ce{CO+} & \ce{CO2} & \ce{C4} & \ce{CH2NH} &
  \ce{CH3CN} & \ce{CH3CHO} & \ce{CH3C3N} & \ce{CH2CCCCH2} &
  \ce{C3H8} & \ce{C6H12} \\
 \ce{P} & \ce{CP} & \ce{H2O} & \ce{CH2O} & \ce{CH2OH} &
  \ce{CH3NC} & \ce{CH3CHS} & \ce{CH3COCN} & \ce{CH3C4H} &
  \ce{C4H10} & \ce{C6H13} \\
  & \ce{CS} & \ce{H2O+} & \ce{CH3} & \ce{CH2PH} &
  \ce{CH3NH} & \ce{CH3NCO} & \ce{CH3OCH2} & \ce{CH3CONH2} &
  \ce{C4H7} & \ce{C6H14} \\
  & \ce{H2} & \ce{H2S} & \ce{CHNO} & \ce{CH2SH} &
  \ce{CH3OH} & \ce{CH3NCO} & \ce{HC6H} & \ce{CH3NHCHO} &
  \ce{C4H8} & \ce{C6H9} \\
  & \ce{HF} & \ce{H3+} & \ce{CO2H+} & \ce{CH3O} &
  \ce{H2C3O} & \ce{CH3NH2} & \ce{HCOOCH3} & \ce{CH3OCH3} &
  \ce{C5H5} & \ce{C7H10} \\
  & \ce{HS} & \ce{HCN} & \ce{H2CN} & \ce{CH3S} &
  \ce{H2CCNH} & \ce{H2CCHOH} & \ce{NC6N} & \ce{HC7N} &
  \ce{C5H6} & \ce{C7H8} \\
  & \ce{N2} & \ce{HCO} & \ce{H2CS} & \ce{CH4} &
  \ce{HC2CHO} & \ce{HC5N} & \ce{NH2CH2CN} &  & \ce{C5H7} &
  \\
  & \ce{NH} & \ce{HCO+} & \ce{H2O2} & \ce{H2CCN} &
  \ce{HC3NH+} &  &  &  & \ce{C5H8} &  \\
  & \ce{NO} & \ce{HNC} & \ce{H3O+} & \ce{H2CCS} &
  \ce{HC4H} &  &  &  & \ce{C5H9} &  \\
  & \ce{NO+} & \ce{HNO} & \ce{HC2O} & \ce{H2NCO+} &
  \ce{HC4N} &  &  &  & \ce{C6H4} &  \\
  & \ce{NS} & \ce{HO2} & \ce{HCCN} & \ce{HC2NC} &
  \ce{HCOCHO} &  &  &  & \ce{C6H5CN} &  \\
  & \ce{O2} & \ce{HOC+} & \ce{HCNH+} & \ce{HC3N} &
  \ce{HNCHCN} &  &  &  & \ce{C6H6} &  \\
  & \ce{OH} & \ce{HPO} & \ce{HCNO} & \ce{HC3O} &
  \ce{HNCHSH} &  &  &  & \ce{C6H7} &  \\
  & \ce{OH+} & \ce{N2H+} & \ce{HCNS} & \ce{HCCCO} &
  \ce{N2H4} &  &  &  & \ce{C6H8} &  \\
  & \ce{PH} & \ce{N2O} & \ce{HNCO} & \ce{HCNCC} &
  \ce{NC4N} &  &  &  & \ce{C7H4} &  \\
  & \ce{PN} & \ce{NH2} & \ce{HNCS} & \ce{HCOOH} &
  \ce{NH2CHO} &  &  &  & \ce{C7H7} &  \\
  & \ce{PO} & \ce{NO2} & \ce{HOCO} & \ce{HNC3} &
  \ce{NH2CHS} &  &  &  & \ce{C8H2} &  \\
  & \ce{S2} & \ce{O3} & \ce{HSCN} & \ce{N2H3} &  &  &
   &  & \ce{C8H6} &  \\
  & \ce{SO} & \ce{OCN} & \ce{HSSH} & \ce{NCCNH+} &  &
   &  &  & \ce{CH2CHCHCH2} &  \\
  &  & \ce{OCS} & \ce{NCSH} & \ce{NH2CN} &  &  &  &
   & \ce{CH2OHCH2OH} &  \\
  &  & \ce{PH2} & \ce{NH3} & \ce{NH4+} &  &  &  &  &
  \ce{CH3C5N} &  \\
  &  & \ce{SO2} & \ce{NO3} & \ce{NHCNH} &  &  &  &
   & \ce{CH3C6H} &  \\
  &  &  & \ce{SO3} &  &  &  &  &  & \ce{CH3COCH3} &
  \\
  &  &  &  &  &  &  &  &  & \ce{CH3COOCH3} &  \\
  &  &  &  &  &  &  &  &  & \ce{CH3OCH2OH} &  \\
  &  &  &  &  &  &  &  &  & \ce{HC11N} &  \\
  &  &  &  &  &  &  &  &  & \ce{HC9N} &  \\
  &  &  &  &  &  &  &  &  & \ce{HCOOC2H5} \\
  \hline\hline
    \end{tabular}
\end{sidewaystable*}
\subsection{Ionization potentials, electron binding energies, and electron kinetic energies}
\begin{figure*}
    \centering
    \includegraphics[width=\textwidth]{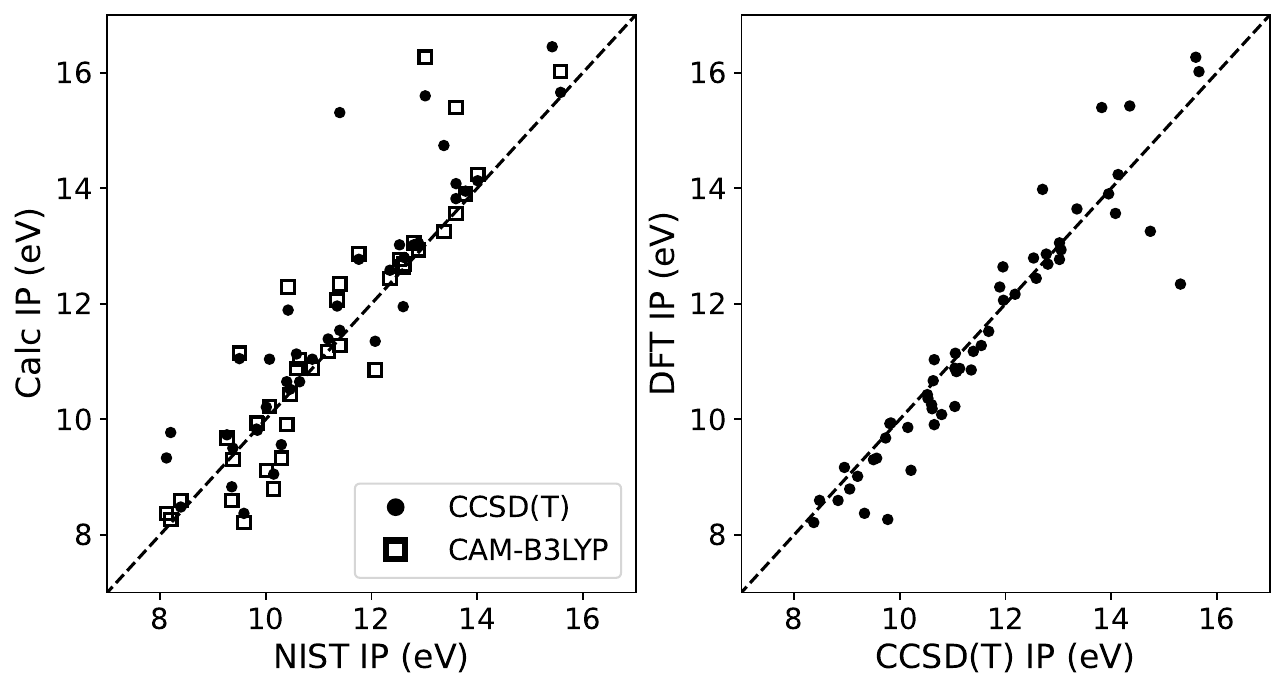}
    \caption{\label{fig:ipcomp}Comparison of ionization potentials presented here against the NIST database. Left: NIST database vs CCSD(T) and CAM-B3LYP DFT calculations (solid and empty square, respectively). Right: CCSD(T) vs CAM-B3LYP DFT ionization potentials. The dashed line represents the one-to-one ratio.}
\end{figure*}

We present here the ionization potentials computed at both the DFT (CAM-B3LYP) and CCSD(T)/CBS levels, and compare them to the experimental value recommended in NIST. Table \ref{tab:ips} shows the resulting ionization potentials in eV for the molecules. Figure \ref{fig:ipcomp} shows a comparison between the different calculations and the NIST database. In general, for molecules with fewer than seven atoms, our calculations are 
consistent with values presented in the NIST database. For molecules with higher ionization potentials, our calculated IPs are generally higher. We find relatively good agreement between the CAM-B3LYP and CCSD(T) calculations, however, there is a general trend that CAM-B3LYP produces smaller IPs compared to CCSD(T) for molecules with IP < 10 eV.

\longtab{
\begin{longtable}{lccc}
\caption{\label{tab:ips} Ionization potentials in eV, computed at the CCSD(T)/CBS, CAM-B3LYP/aug-cc-pVQZ  and from NIST.}\\
\toprule\toprule
Molecule & IP (CCSD(T)) & IP (DFT) & IP\footnote{\url{https://webbook.nist.gov/chemistry/ie-ser/}} (NIST) \\
\midrule
\endfirsthead
\caption{continued.}\\
\toprule\toprule
Molecule & IP (CCSD(T)) & IP (DFT) & IP (NIST)  \\
\midrule
\endhead
\bottomrule
\endfoot
Diatomic & IP (CCSD(T)) & IP (DFT) & IP (NIST)  \\
\hline
\ce{C2} & 15.31 & 12.34 & 11.40 \\
\ce{CH} & 10.65 & 11.03 & 10.64 \\
\ce{CN} & 13.82 & 15.40 & 13.60 \\
\ce{CO} & 14.13 & 14.24 & 14.01 \\
\ce{CP} &  & 11.79 &  \\
\ce{CS} &  & 11.45 &  \\
\ce{H2} & 16.45 &  & 15.42 \\
\ce{HS} & 11.89 & 12.29 & 10.42 \\
\ce{N2} & 15.66 & 16.02 & 15.58 \\
\ce{NH} & 11.68 & 11.52 &  \\
\ce{NO} & 9.73 & 9.68 & 9.26 \\
\ce{NS} & 8.95 & 9.17 &  \\
\ce{O2} & 11.35 & 10.85 & 12.07 \\
\ce{OH} & 15.60 & 16.27 & 13.02 \\
\ce{PH} & 9.05 & 8.79 & 10.15 \\
\ce{PN} &  & 11.89 &  \\
\ce{PO} & 8.48 & 8.59 & 8.39 \\
\ce{S2} & 8.83 & 8.59 & 9.36 \\
\ce{SO} & 9.56 & 9.33 & 10.29 \\
\hline
Triatomic & IP (CCSD(T)) & IP (DFT) & IP (NIST)  \\
\hline
\ce{C3} & 11.95 & 12.64 & 12.60 \\
\ce{CCN} &  & 11.21 &  \\
\ce{CCO} & 10.60 & 10.25 &  \\
\ce{CH2} & 10.65 & 9.91 & 10.40 \\
\ce{CNO} & 12.77 & 12.86 & 11.76 \\
\ce{CO2} & 13.95 & 13.90 & 13.78 \\
\ce{H2O} & 12.80 & 12.69 & 12.62 \\
\ce{H2S} & 10.52 & 10.43 & 10.46 \\
\ce{HCN} & 14.08 & 13.56 & 13.60 \\
\ce{HCO} & 9.33 & 8.37 & 8.12 \\
\ce{HNC} & 12.18 & 12.16 &  \\
\ce{HNO} & 10.61 & 10.18 &  \\
\ce{HPO} & 10.53 & 10.36 &  \\
\ce{N2O} & 13.05 & 12.93 & 12.89 \\
\ce{NH2} & 12.53 & 12.79 &  \\
\ce{NO2} & 8.37 & 8.21 & 9.59 \\
\ce{O2H} & 11.96 & 12.06 & 11.35 \\
\ce{O3} & 13.02 & 12.77 & 12.53 \\
\ce{OCN} & 13.35 & 13.64 &  \\
\ce{OCS} & 11.39 & 11.18 & 11.18 \\
\ce{PH2} & 9.83 & 9.94 & 9.82 \\
\ce{SO2} & 12.58 & 12.44 & 12.35 \\
\hline
4-atomic & IP (CCSD(T)) & IP (DFT) & IP (NIST)  \\
\hline
\ce{C2H2} & 11.54 & 11.28 & 11.40 \\
\ce{C2N2} & 14.74 & 13.25 & 13.37 \\
\ce{C3N} & 14.35 & 15.42 &  \\
\ce{C4} &  & 10.56 &  \\
\ce{CH3} & 9.81 & 9.92 & 9.84 \\
\ce{H2C3O} &  & 10.59 &  \\
\ce{H2CN} & 10.63 & 10.67 &  \\
\ce{H2CO} & 11.04 & 10.88 & 10.88 \\
\ce{H2CS} & 9.50 & 9.30 & 9.38 \\
\ce{HC2N} & 10.79 & 10.08 &  \\
\ce{HC2O} & 11.05 & 11.14 & 9.50 \\
\ce{HCNS} & 9.20 & 9.01 &  \\
\ce{HNCS} & 10.15 & 9.86 &  \\
\ce{HOCO} & 9.77 & 8.27 & 8.20 \\
\ce{HOOH} & 11.13 & 10.88 & 10.58 \\
\ce{HSCN} & 11.07 & 10.83 &  \\
\ce{HSSH} & 10.21 & 9.12 & 10.01 \\
\ce{NCSH} & 11.07 & 10.83 &  \\
\ce{NH3} & 11.04 & 10.22 & 10.07 \\
\ce{NO3} & 12.70 & 13.98 &  \\
\ce{SO3} & 13.02 & 13.05 & 12.80 \\
\hline
5-atomic & IP (CCSD(T)) & IP (DFT) & IP (NIST)  \\
\hline
\ce{C2H3} &  & 8.73 &  \\
\ce{C2H3CHO} &  & 9.99 &  \\
\ce{C4H} &  & 14.39 &  \\
\ce{CH2NH} &  & 9.92 &  \\
\ce{CH2OH} &  & 7.62 &  \\
\ce{CH2PH} &  & 10.04 &  \\
\ce{CH2SH} &  & 7.63 &  \\
\ce{CH3O} &  & 8.89 &  \\
\ce{CH3S} &  & 10.24 &  \\
\ce{CH4} &  & 12.72 &  \\
\ce{H2CCN} &  & 10.42 &  \\
\ce{H2CCS} &  & 8.86 &  \\
\ce{HC3N} &  & 11.48 &  \\
\ce{HC3O} &  & 11.26 &  \\
\ce{HCCCO} &  & 7.57 &  \\
\ce{HCCNC} &  & 11.04 &  \\
\ce{HCNCC} &  & 11.09 &  \\
\ce{N2H3} &  & 7.81 &  \\
\hline
6-atomic & IP (CCSD(T)) & IP (DFT) & IP (NIST)  \\
\hline
\ce{C2H4} &  & 10.40 &  \\
\ce{C3H3} &  & 8.76 &  \\
\ce{CH2CCO} &  & 9.02 &  \\
\ce{CH2NH2} &  & 6.28 &  \\
\ce{CH3NCO} &  & 9.59 &  \\
\ce{CH3NH} &  & 15.09 &  \\
\ce{CH3OH} &  & 10.80 &  \\
\ce{HCOCHO} &  & 10.16 &  \\
\ce{HNCHSH} &  & 9.71 &  \\
\ce{N2H4} &  & 6.82 &  \\
\ce{NC4N} &  & 11.67 &  \\
\ce{NH2CHO} &  & 10.13 &  \\
\ce{NH2CHS} &  & 8.54 &  \\
\ce{c-H2C3O} &  & 9.45 &  \\
\ce{l-HC4H} &  & 9.96 &  \\
\hline
7-atomic & IP (CCSD(T)) & IP (DFT) & IP (NIST)  \\
\hline
\ce{C2H5} &  & 8.19 &  \\
\ce{CH2CCH2} &  & 7.24 &  \\
\ce{CH2CHCN} &  & 10.72 &  \\
\ce{CH3CHS} &  & 8.85 &  \\
\ce{CH3NH2} &  & 9.00 &  \\
\hline
8-atomic & IP (CCSD(T)) & IP (DFT) & IP (NIST)  \\
\hline
\ce{C2H3NH2} &  & 7.98 &  \\
\ce{C2H6} &  & 11.59 &  \\
\ce{C3H5} &  & 7.77 &  \\
\ce{C4H4} &  & 9.32 &  \\
\ce{CH3C3N} &  & 10.59 &  \\
\ce{CH3COCN} &  & 11.16 &  \\
\ce{CH3OCH2} &  & 7.05 &  \\
\ce{HCOOCH3} &  & 10.74 &  \\
\ce{HNCCC} &  & 9.80 &  \\
\ce{NC6N} &  & 10.69 &  \\
\ce{l-HC6H} &  & 9.26 &  \\
\hline
9-atomic & IP (CCSD(T)) & IP (DFT) & IP (NIST)  \\
\hline
\ce{C4H5} &  & 7.96 &  \\
\ce{C5H4} &  & 10.03 &  \\
\ce{C8H} &  & 8.81 &  \\
\ce{CH2CCCCH2} &  & 8.55 &  \\
\ce{CH3C4H} &  & 9.26 &  \\
\ce{CH3OCH3} &  & 9.85 &  \\
\ce{l-C3H6} &  & 9.57 &  \\
\hline
10-atomic and more & IP (CCSD(T)) & IP (DFT) & IP (NIST)  \\
\hline
\ce{C10H2} &  & 8.50 &  \\
\ce{C2H5CHO} &  & 9.92 &  \\
\ce{C3H5CN} &  & 10.00 &  \\
\ce{C3H7} &  & 10.99 &  \\
\ce{C3H8} &  & 10.99 &  \\
\ce{C4H10} &  & 10.89 &  \\
\ce{C4H7} &  & 7.58 &  \\
\ce{C4H8} &  & 10.89 &  \\
\ce{C5H5} &  & 7.80 &  \\
\ce{C5H6} &  & 8.74 &  \\
\ce{C5H7} &  & 7.42 &  \\
\ce{C5H8} &  & 8.35 &  \\
\ce{C6H4} &  & 8.74 &  \\
\ce{C6H7} &  & 7.31 &  \\
\ce{C6H8} &  & 8.70 &  \\
\ce{C7H4} &  & 7.98 &  \\
\ce{C8H2} &  & 8.81 &  \\
\ce{C8H6} &  & 8.65 &  \\
\ce{CH2CHCHCH2} &  & 8.80 &  \\
\ce{CH3C6H} &  & 8.73 &  \\
\ce{CH3CCH} &  & 10.21 &  \\
\ce{CH3CH2OH} &  & 10.38 &  \\
\ce{CH3COCH3} &  & 9.64 &  \\
\ce{c-C6H6} &  & 9.17 &  \\
\ce{C14H10} &  & 7.21 &  \\
\ce{C5H12} &  & 10.76 &  \\
\ce{C6H10} &  & 8.27 &  \\
\ce{C6H12} &  & 9.29 &  \\
\ce{C6H14} &  & 10.77 &  \\
\ce{C6H9} &  & 6.97 &  \\
\ce{C7H8} &  & 9.09 &  \\
\end{longtable}
}

As part of the database, we also present the molecular orbital binding and kinetic energies for each molecule. An example result of this, for CO and \ce{H2O}, is presented in Table \ref{tab:coOrb} and Table \ref{tab:h2oOrb} respectively. The tables show both, the eBEs computed from the canonical HF orbitals and from the EPT theory. The eBE computed with the latter method are smaller than the ones obtained with HF since they are corrected to include the instantaneous electron-electron repulsion (electron correlation) in its energy, which increases the energy of the orbitals thus making it easier for an electron to detach. According to Koopmans' theorem, the lowest eBE is equal to the ionization potential of the molecule. Comparing the the values of the EPT-eBE and the IP computed at our best level of theory (CCSD(T)/CBS) excellent agreement can be seen with discrepancies of around 0.2 eV and much improved with respect to HF-eBEs. Given that the lowest eBE carries the most weight in the BEB cross section, ensuring its accuracy is paramount. Our results not only exhibit a marked improvement over canonical HF values but also suggest that the error related to the electronic structure is minimal.  However, for the deep-lying orbitals, no reliable EPR-eBEs can be computed thus we chose to use the HF values for these orbitals. 

We find good agreement between our HF-BE data and the NIST electron-impact cross-section database, which uses data from \citet{Hwang1996} computed at the HF/6-311-G level for the eBE. This is expected as the only difference between the NIST and our HF orbital binding energies is a larger atomic orbital  basis set in the latter case. However the incorporation of EPT-BE into our database enhances the overall data quality compared to the existing NIST data. We thus conclude that our calculations of orbital binding energies are derived from a robust theoretical framework, offering the required level of accuracy given the underlying assumptions in our cross section calculations.

\begin{table}
    \centering
    \caption{\label{tab:coOrb} Molecular orbitals of \ce{CO}, computed at the MP2/aug-cc-pVTZ level. All energies are in eV.  The orbital kinetic energy (KE), the orbital binding energy based on canonical HF orbitals (BE-HF), the orbital binding energy based on EPT computation (BE-EPT) and the corresponding pole strength (PS) are displayed. *** denote no data, since EPT energies for pole strengths less than 0.8 ( electrons deep within the potential) are not considered.}
    \begin{tabular}{cr|rr|r}
    \hline
    Orbital number & KE & BE-HF & BE-EPT & PS \\
    \hline 
    1 & 794.32671 & 562.41699 & *** & *** \\
    2 & 436.29414 & 309.24685 & *** & *** \\
    3 & 78.26264 & 41.33263 & *** & *** \\
    4 & 71.85773 & 21.8887 & *** & *** \\
    5 & 53.94510 & 17.35920 & 17.015 & 0.903 \\
    6 & 53.94510 & 17.35920 & 17.015 & 0.903 \\
    7 & 42.78816 & 15.11922 & 14.266 & 0.910 \\
    \hline
    \end{tabular}
\end{table}

\begin{table}
    \centering
    \caption{\label{tab:h2oOrb} Same as Table \ref{tab:coOrb} but for \ce{H2O}.}
    \begin{tabular}{cr|rr|r}
    \hline
    Orbital number & KE & BE-HF & BE-EPT & PS \\
    \hline 
    1 & 794.31517 & 559.67367 & *** & *** \\
    2 & 71.17519 & 36.81820 & *** & *** \\
    3 & 48.69143 & 19.49103 & 18.872 & 0.933 \\
    4 & 58.43876 & 15.92808 & 14.802 & 0.925 \\
    5 & 60.77375 & 13.88698 & 12.554 & 0.921 \\
    \hline
    \end{tabular}
\end{table}

Finally, we also include the optimized geometries at the MP2/aug-cc-pVTZ and DF-CCSD(T)-F12/cc-pVDZ-F12 levels presented here and the sample from \citet{Heathcote2018} in the database. The geometries are not appreciably different from each other. All geometries were checked that represent minimum energy structures by ensuring that no imaginary frequencies in the diagonalized Hessian matrix is present.

\subsection{Low-energy electron cross sections}
As described earlier, we have computed the cross sections using three different approaches: the screening-corrected additivity rule (SCAR), the binary-encounter Bethe (BEB) model, and the new damped BEB cross section (dBEB). 

\begin{figure*}
    \centering
    \includegraphics[width=\textwidth]{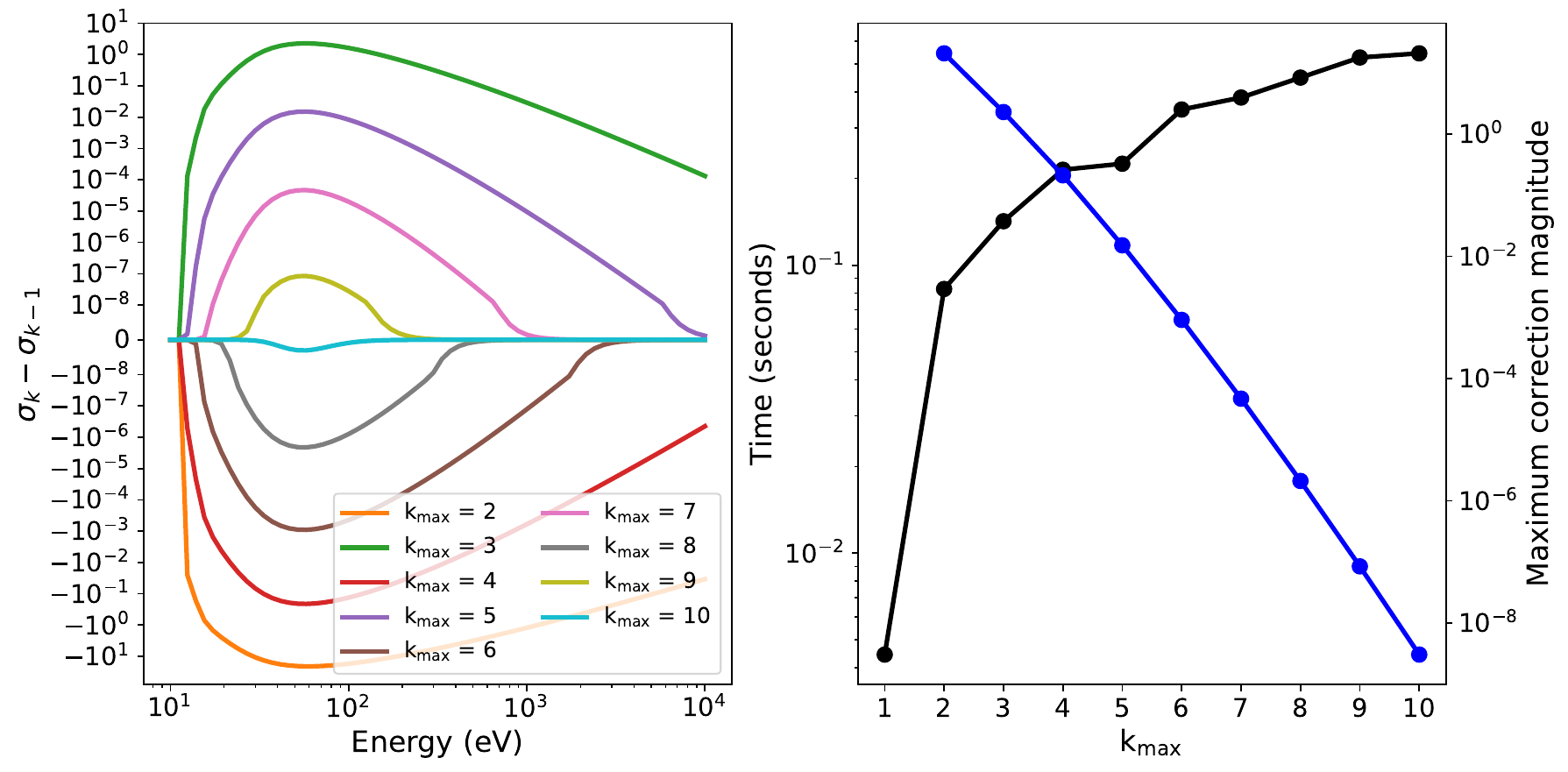}
    \caption{\label{fig:benchmark}Left: k-atom screening corrections as a function of electron energy for \ce{C5H10}. Right: Time to compute $\sigma_{\rm SCAR}(E_e)$ for a given maximum k-atom screening correction (black) and the magnitude of the maximal correction (blue) as a function of the maximum $k$.}
\end{figure*}

The SCAR method is inherently recursive, so care must be taken to avoid exponential increases in computational time with the number of atoms. We utilized recursive caching to speed up the calculations, and only computed to a maximum of the 10-atom screening correction. Figure \ref{fig:benchmark} shows the individual screening corrections, the magnitude of the correction and the total time as a function of the maximum screening terms kept. We find that the screening terms past the eight-atom screening correction are negligible contributions to the total cross section, which is dominated by the first few correction factors.

\begin{figure}
    \centering
    \includegraphics[width=0.5\textwidth]{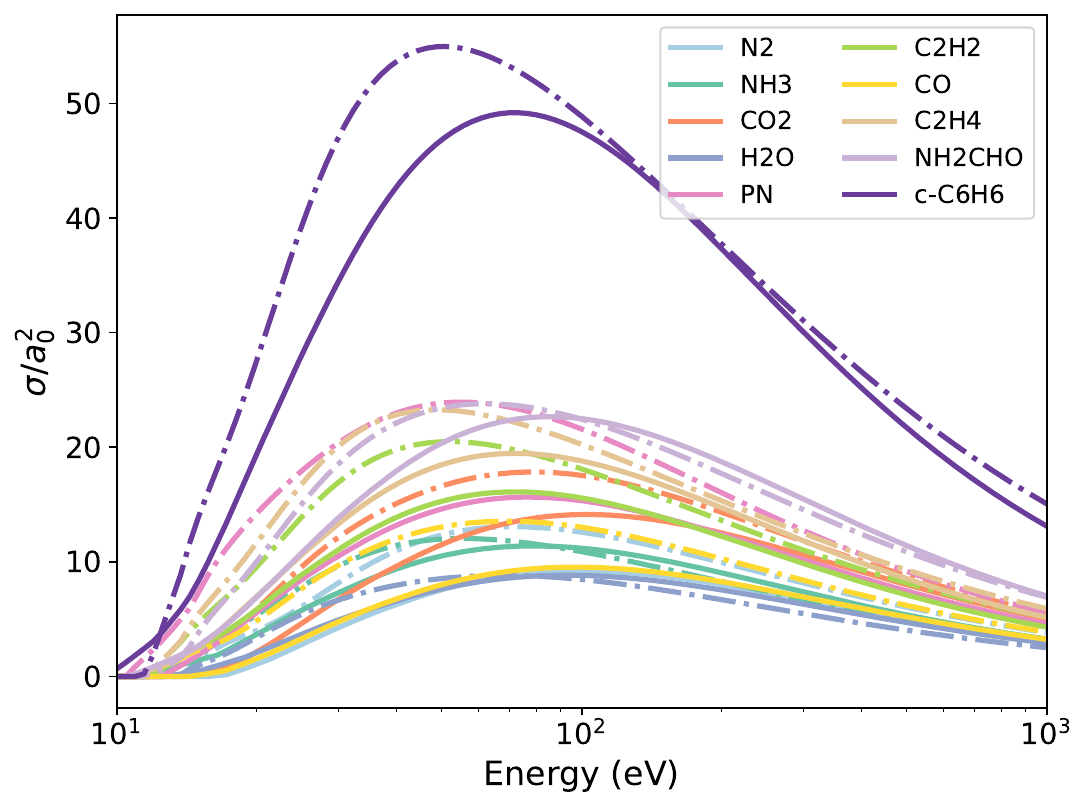}
    \caption{\label{fig:xsec_sample}Sample of the total electron ionization cross sections for a sub-sample of the molecules. Solid denotes cross sections calculated using the BEB method, and the dotted line uses the SCAR rule.}
\end{figure}

We show results for a subset of the molecules, ranging from simple to complex, to compare the difference between the SCAR and BEB cross sections. Figure \ref{fig:xsec_sample} shows these for a subsample of 10 molecules, ranging in size from \ce{CO} to \ce{c-C6H6}. We find that the SCAR approach generally overestimates the cross section, although the impact is most pronounced in \ce{c-C6H6} due to the overlapping of the molecular orbitals. However, given the speed of computing these cross sections, they may be useful for first investigations if the optimized molecule geometry is known {\it a priori}. While we encourage the use of BEB cross sections, we include the database a python script to compute the SCAR cross section from a provided optimized geometry.

\begin{figure}
    \centering
    \includegraphics[width=0.5\textwidth]{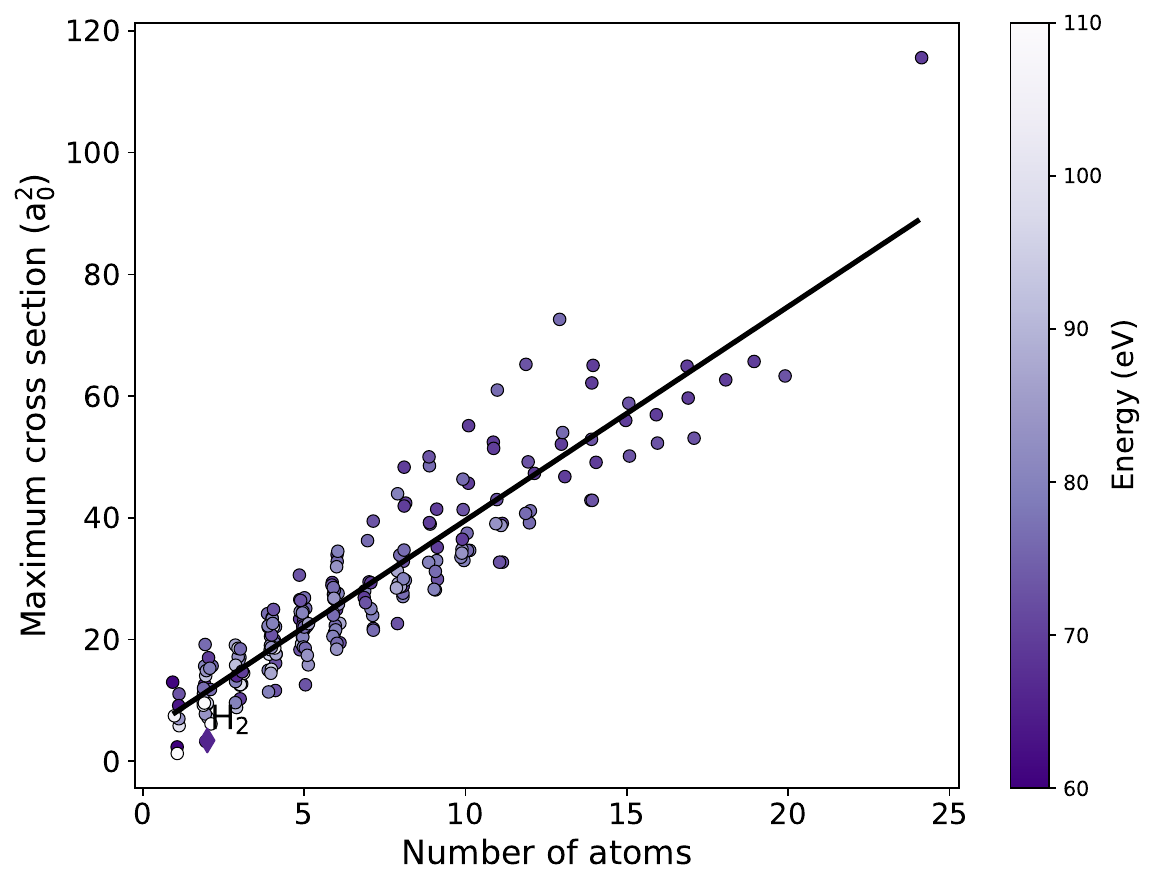}
    \caption{\label{fig:maxXS}Maximum of the electron-impact ionization cross section as a function of the number of atoms in a molecule. The color denotes the electron energy in eV the maximum occurs.}
\end{figure}

\begin{figure*}
    \centering
    \begin{subfigure}[b]{0.3\textwidth}
        \centering
        \includegraphics[width=\textwidth]{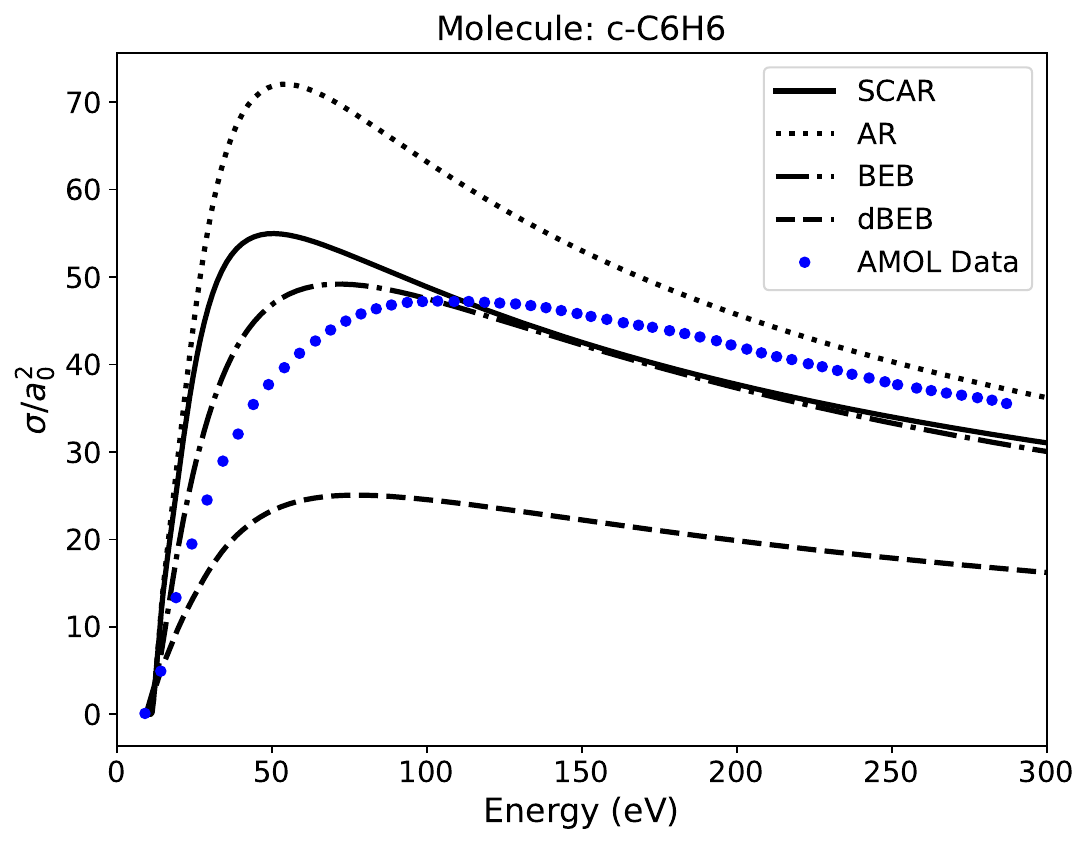}
        \caption{\ce{c-C6H6}}
    \end{subfigure}
    \begin{subfigure}[b]{0.3\textwidth}
        \centering
        \includegraphics[width=\textwidth]{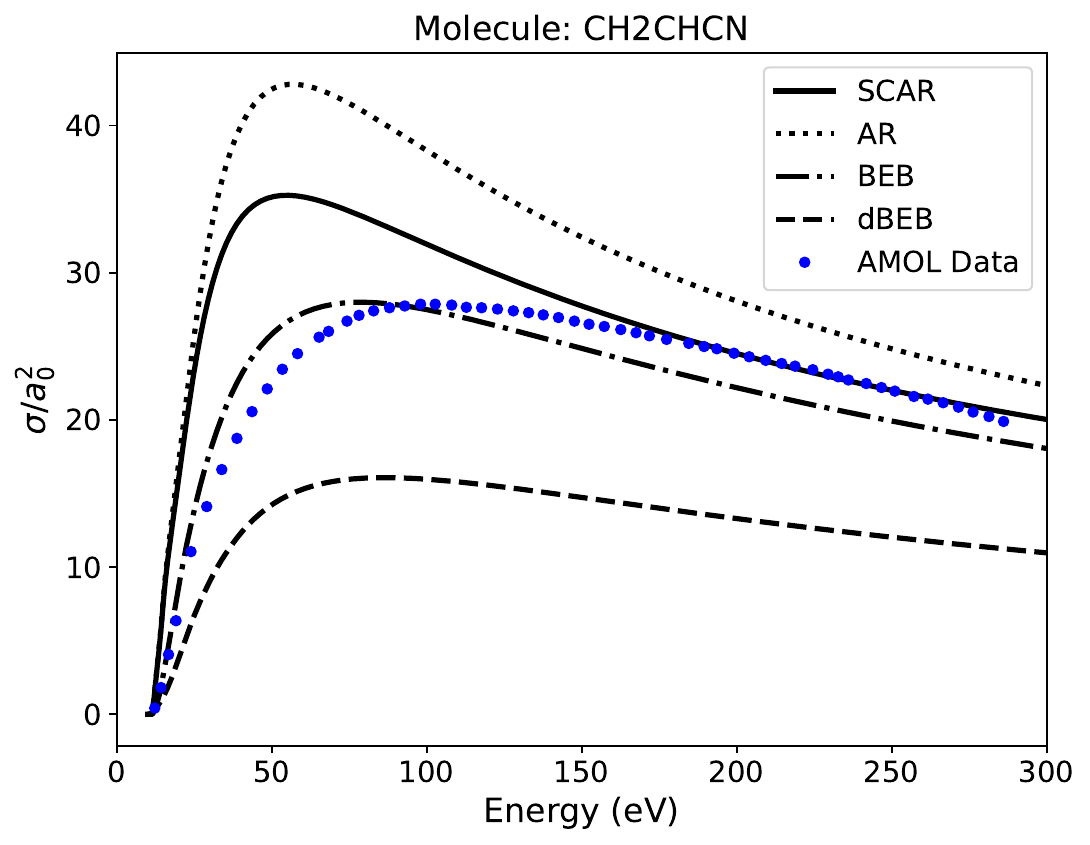}
        \caption{\ce{CH2CHCN}}
    \end{subfigure}
    \begin{subfigure}[b]{0.3\textwidth}
        \centering
        \includegraphics[width=\textwidth]{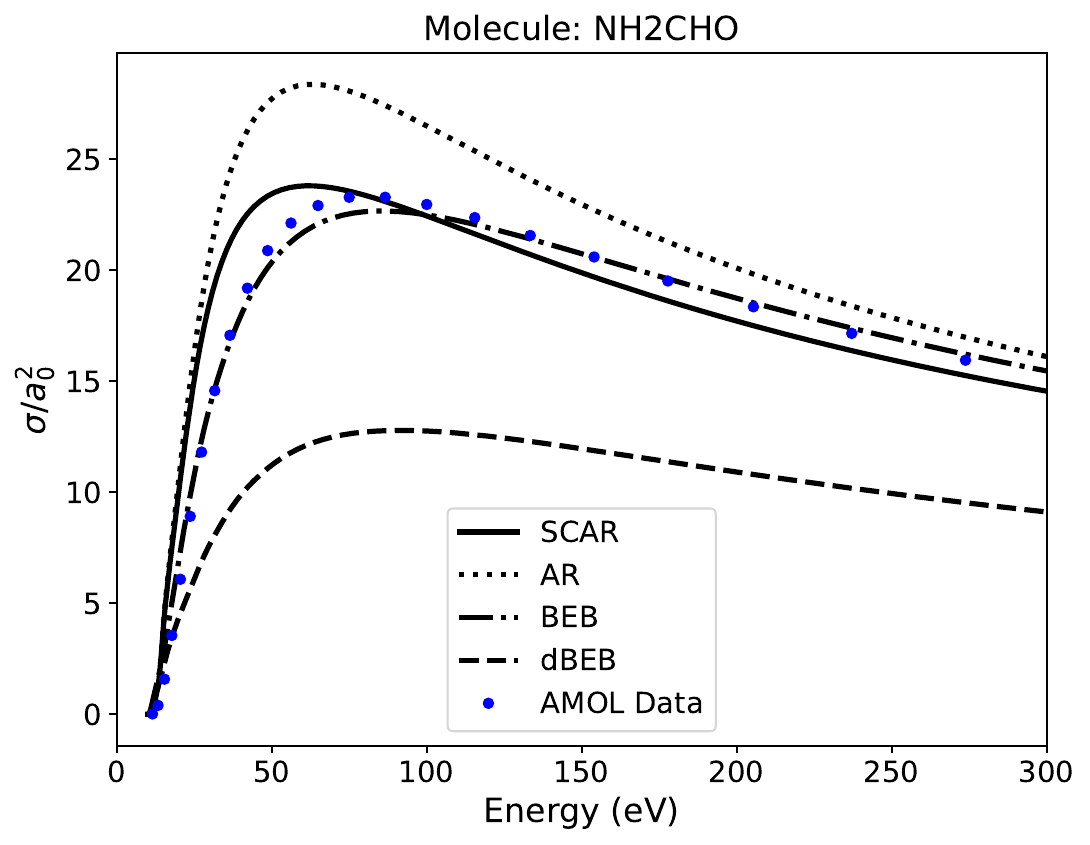}
        \caption{\ce{NH2CHO}}
    \end{subfigure}
    \begin{subfigure}[b]{0.3\textwidth}
        \centering
        \includegraphics[width=\textwidth]{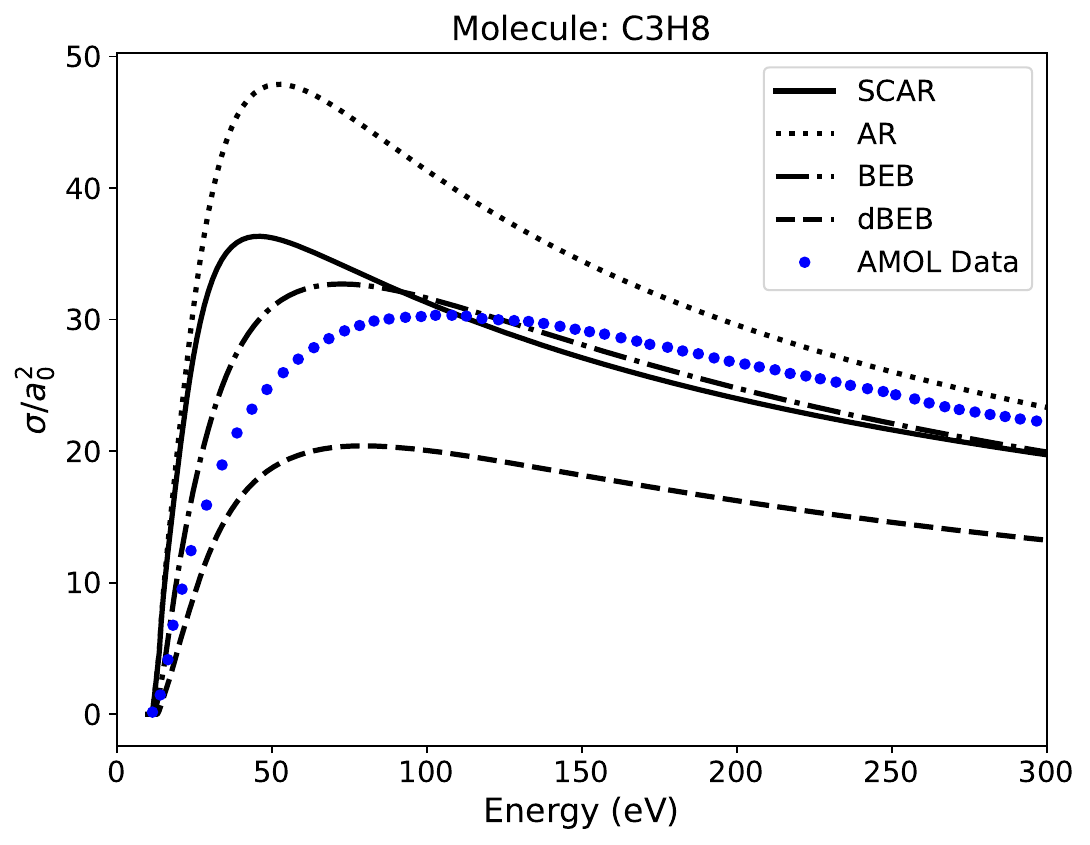}
        \caption{\ce{C3H8}}
    \end{subfigure}
    \begin{subfigure}[b]{0.3\textwidth}
        \centering
        \includegraphics[width=\textwidth]{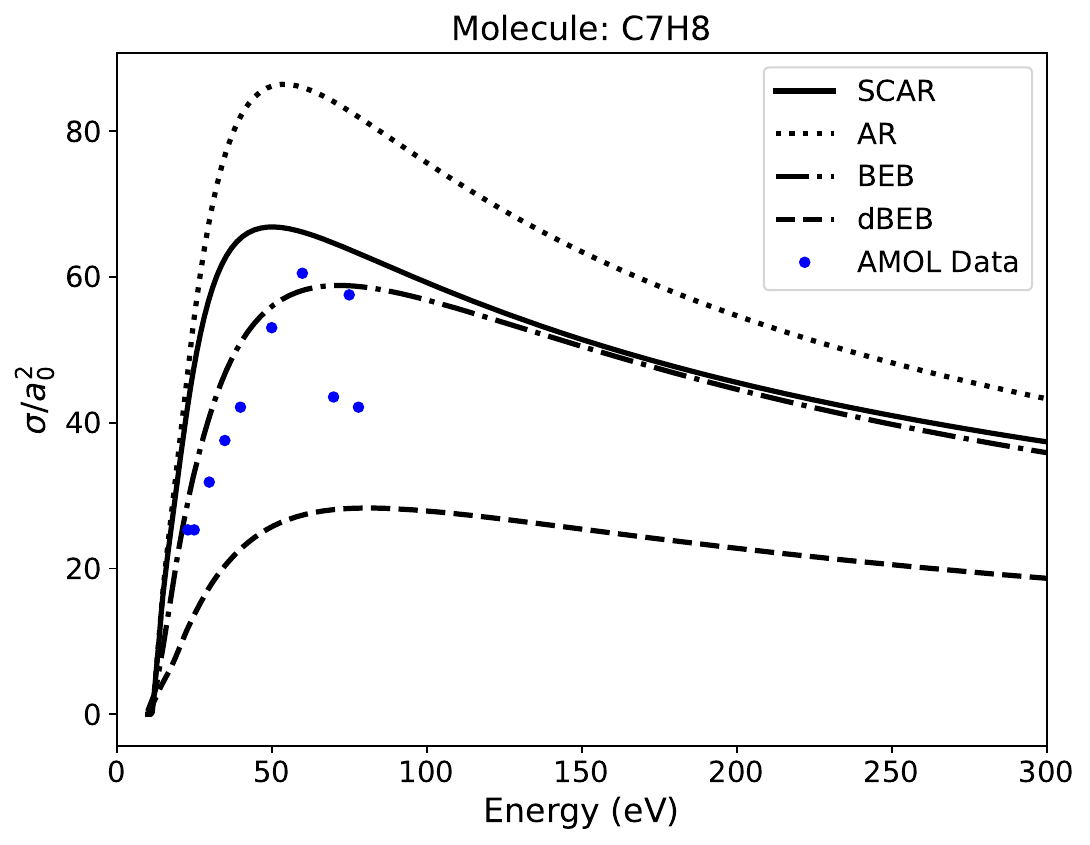}
        \caption{\ce{C7H8}}
    \end{subfigure}
    \begin{subfigure}[b]{0.3\textwidth}
        \centering
        \includegraphics[width=\textwidth]{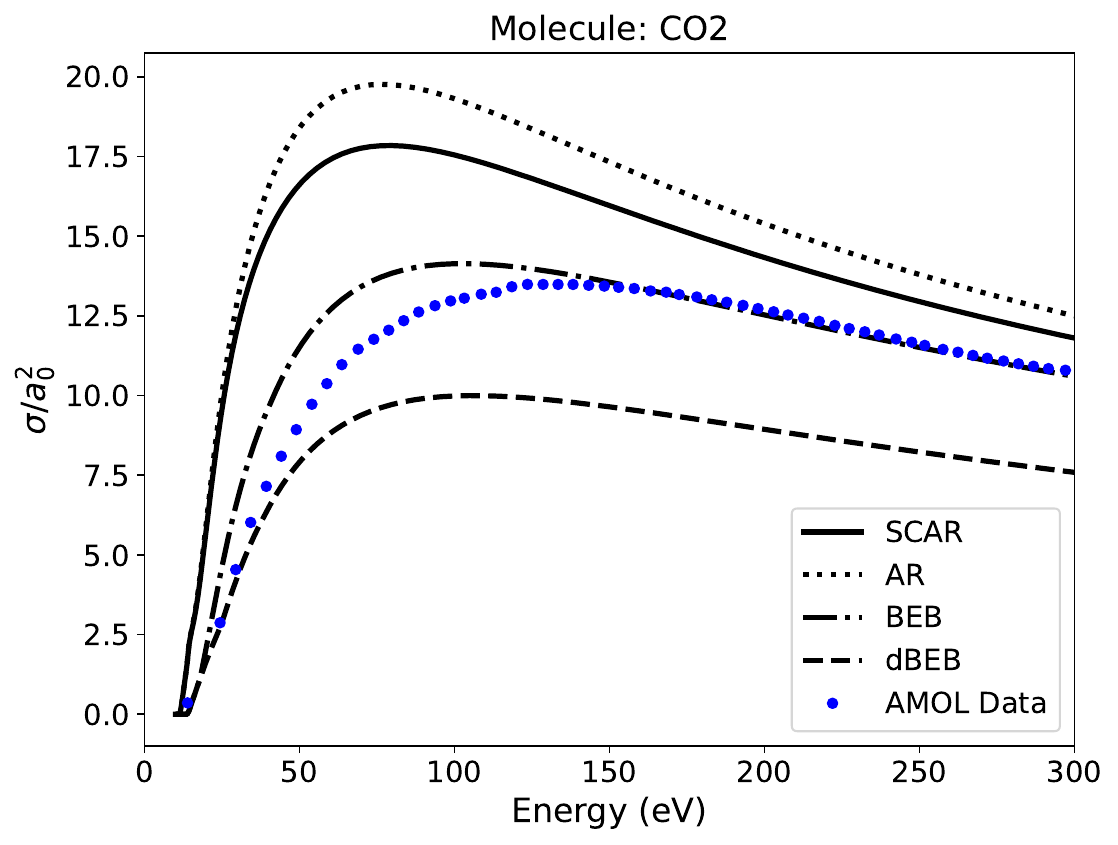}
        \caption{\ce{CO2}}
    \end{subfigure}
    \caption{\label{fig:elecCrossSamp} Electron ionization cross sections computed with the AR (dashed), SCAR (solid), and BEB (dashed-dotted) methods compared against experimental data from the NIFS database (blue dots).}
\end{figure*}

We perform a closer investigation of the model predictions for a subset of these molecules for which experimental data is available. Figure~\ref{fig:elecCrossSamp} shows a comparison with experimental data for a subset of molecules, including the simple molecule \ce{CO2}, symmetric ring \ce{c-C6H6}, prebiotic species of interest \ce{NH2CHO} and carbon chains. Experimental and theoretical cross-section data are known to deviate quite significantly (see discussion in \citealt{Zhou2019}). In particular, experimental cross sections include double ionizations and Auger ionizations, thus overestimating the single ionization cross section. The cross sections we present here are for single ionization events, with multiple ionizations being left for future work. We find that with our computed data, the BEB slightly overesitmates the cross section  at low energy regions while underestimating it in higher energy regions. Nonetheless the values from the BEB calculation are in much better agreement with experimental results than the SCAR or the AR model. Given that the experimental curve should serve as an upper limit for the theoretical BEB model, and noting that even with highly accurate eBEs it still overestimates the cross section, we suggest a modified BEB model in which the the contribution from deeper-lying electron ionizations are scaled, to better balance its weight in the total cross section. As expected, we find that the dBEB corss-section under-predicts the experimental value. Future calculations using the more complicated binary-encounter dipole model \citep{Kim1994} will be investigated for future releases. In our further analysis, we use the BEB cross sections to calculate the reaction rates, since towards higher energies they tend towards better agreement with data.

\subsection{Molecular cloud rate coefficients}
\begin{figure*}
    \centering
    \begin{subfigure}[b]{0.95\textwidth}
        \centering
        \includegraphics[width=\textwidth]{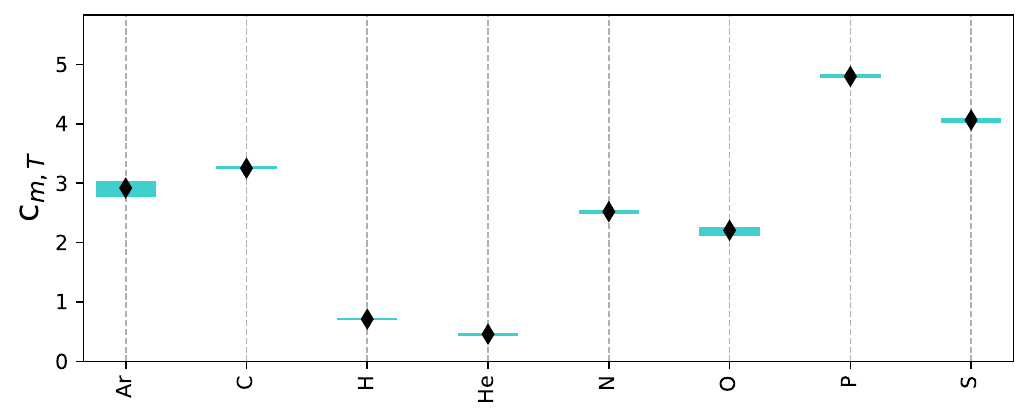}
    \end{subfigure}
    \begin{subfigure}[b]{0.95\textwidth}
        \centering
        \includegraphics[width=\textwidth]{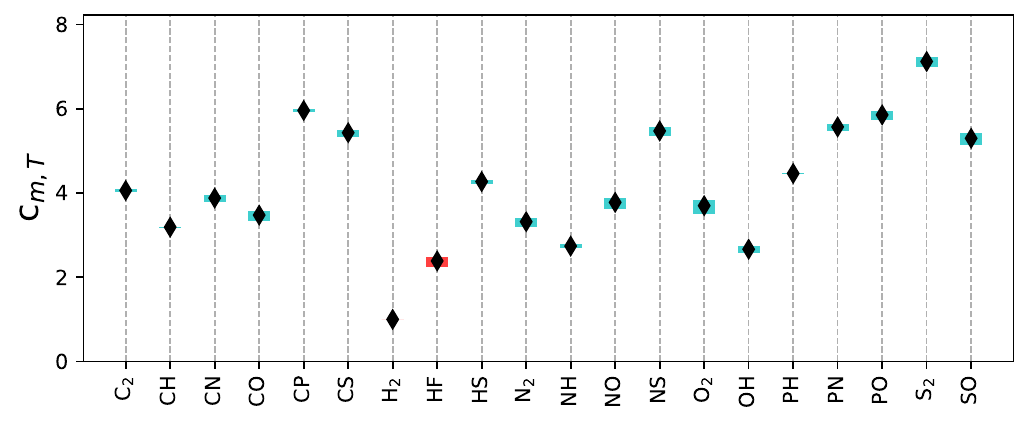}
    \end{subfigure}
    \begin{subfigure}[b]{0.95\textwidth}
        \centering
        \includegraphics[width=\textwidth]{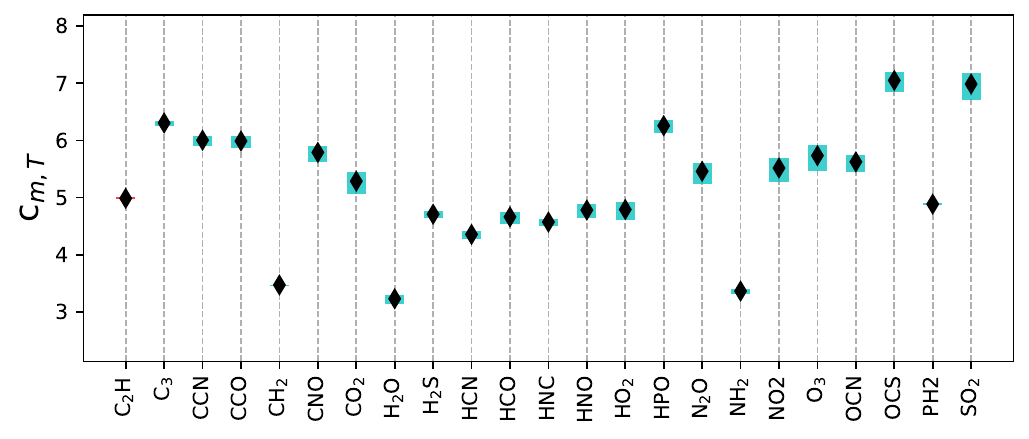}
    \end{subfigure}
    \caption{\label{fig:coeff_1} Reaction rate coefficients, following Equation \ref{eq:coeffs} for all neutral molecules in the database. The boxes show the minimum and maximum coefficients in the column density range $10^{20}$ cm$^{-2}$ $\le N_{\ce{H2}} \le$ $10^{23}$ cm$^{-2}$, plus signs show the mean. The cyan boxes denote molecules using MP2 calculations while red denote calculations from \citet{Heathcote2018}.}
\end{figure*}

\begin{figure*}
    \centering
    \begin{subfigure}[b]{0.95\textwidth}
        \centering
        \includegraphics[width=\textwidth]{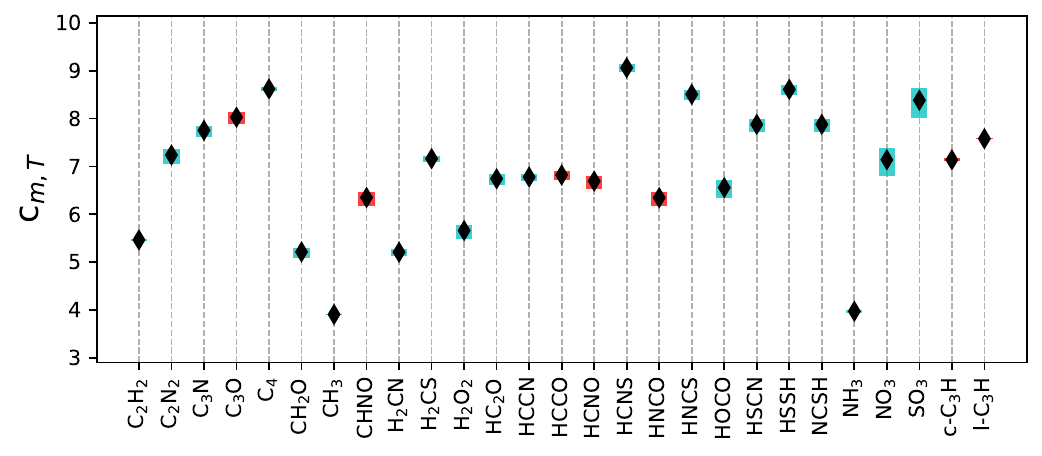}
    \end{subfigure}
    \begin{subfigure}[b]{0.95\textwidth}
        \centering
        \includegraphics[width=\textwidth]{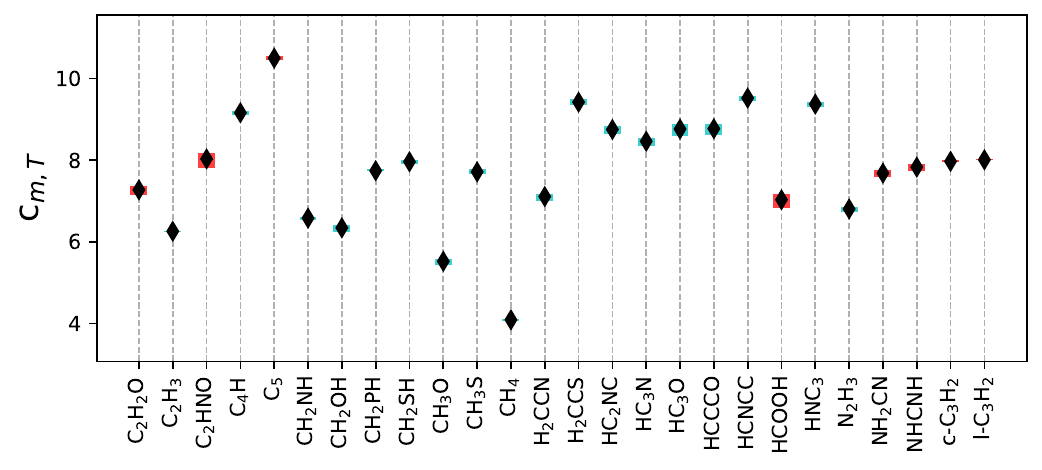}
    \end{subfigure}
    \begin{subfigure}[b]{0.95\textwidth}
        \centering
        \includegraphics[width=\textwidth]{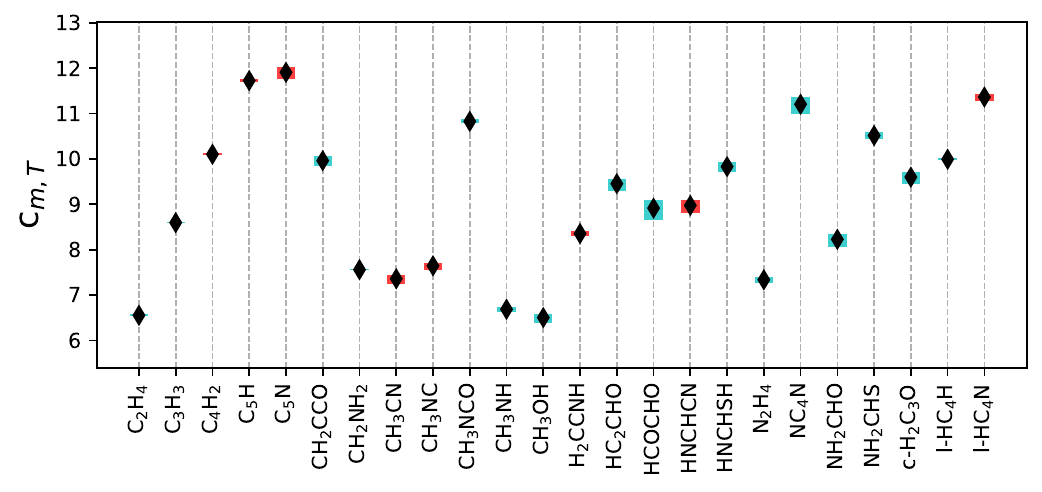}
    \end{subfigure}
    \caption{\label{fig:coeff_2}Figure ~\ref{fig:coeff_1} cont.}
\end{figure*}

\begin{figure*}
    \centering
        \begin{subfigure}[b]{0.95\textwidth}
        \centering
        \includegraphics[width=\textwidth]{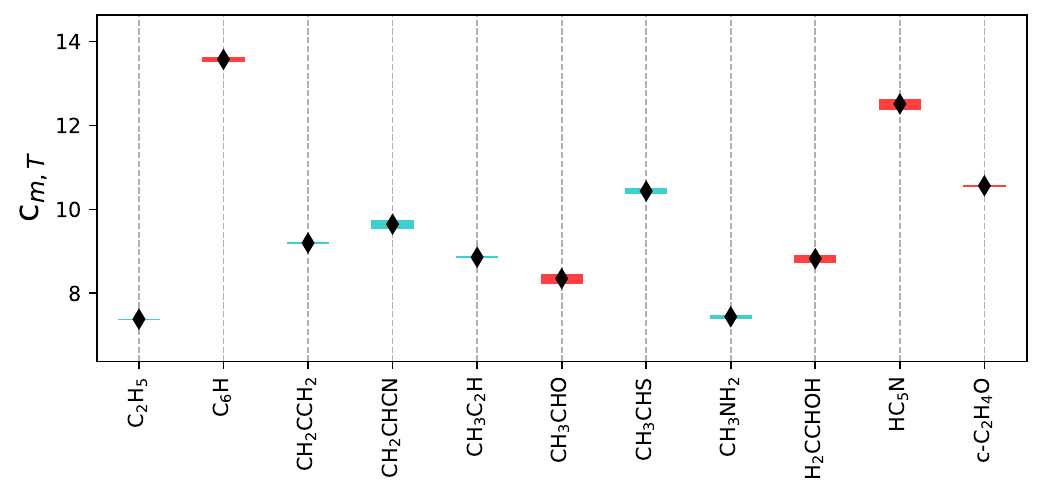}
    \end{subfigure}
    \begin{subfigure}[b]{0.95\textwidth}
        \centering
        \includegraphics[width=\textwidth]{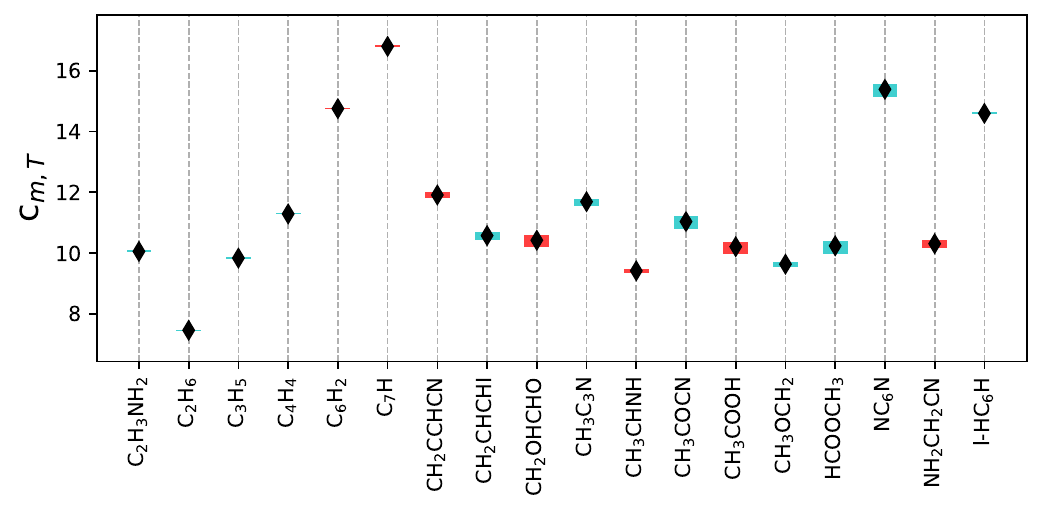}
    \end{subfigure}
    \begin{subfigure}[b]{0.95\textwidth}
        \centering
        \includegraphics[width=\textwidth]{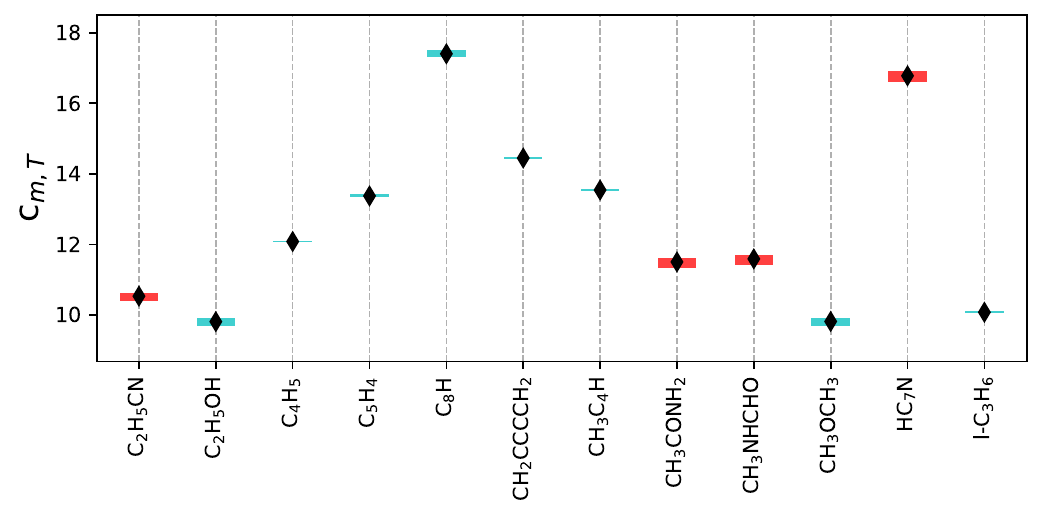}
    \end{subfigure}
    \caption{\label{fig:coeff_3}Figure ~\ref{fig:coeff_2} cont.}
\end{figure*}

\begin{figure*}
    \centering
    \begin{subfigure}[b]{0.95\textwidth}
        \centering
        \includegraphics[width=\textwidth]{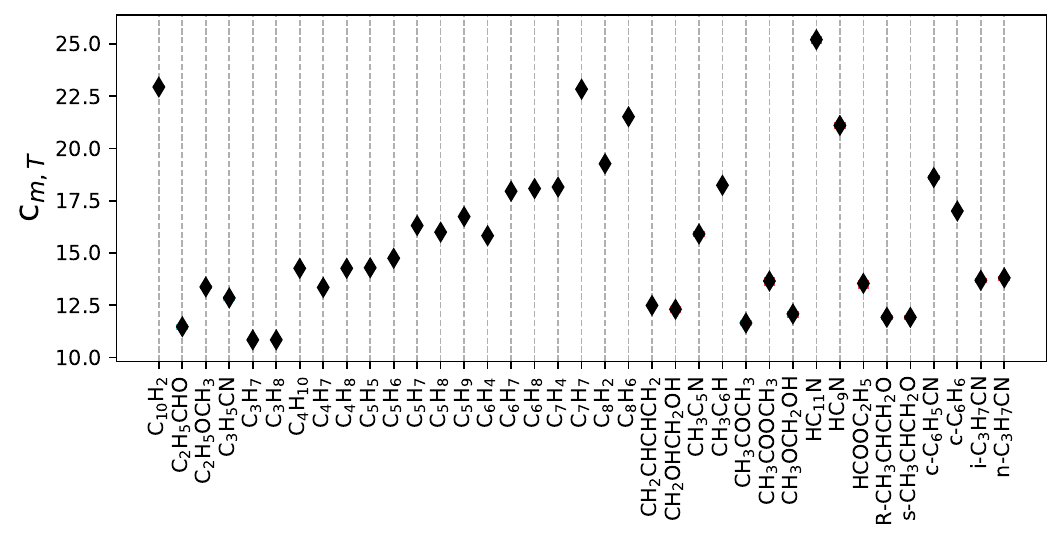}
    \end{subfigure}
    \begin{subfigure}[b]{0.95\textwidth}
        \centering
        \includegraphics[width=\textwidth]{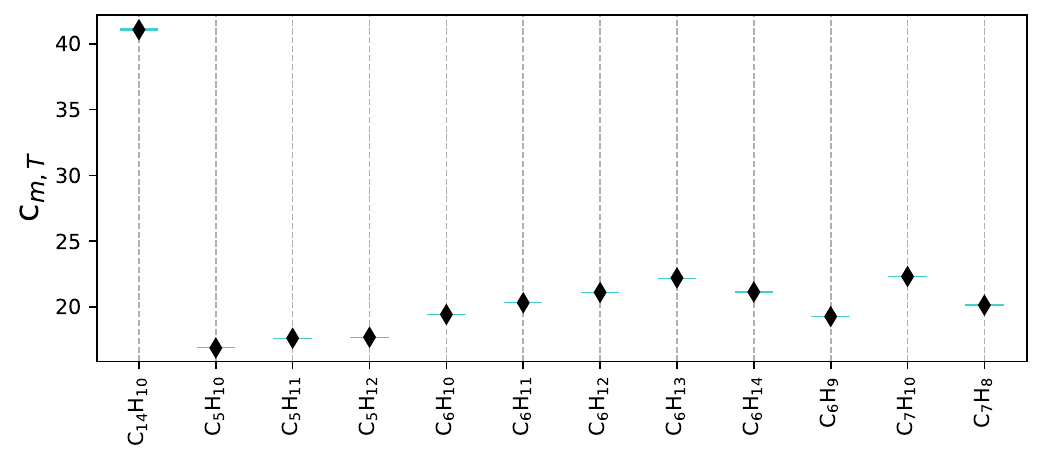}
    \end{subfigure}
    \caption{\label{fig:coeff_4}Figure ~\ref{fig:coeff_3} cont.}
\end{figure*}

We discuss here the molecular cloud reaction rates coefficients, scaled to the total ionization rate (see Section \ref{sec:crion}). Figures \ref{fig:coeff_1} - \ref{fig:coeff_4} show the coefficients, $c_{m,T}$, for our database sample. There is a general trend of an increase in $c_{m,T}$ with the size of the molecule. However, for a given number of atoms, there is still substantial scatter due to the geometry and composition. Our reaction rate is marginally less than that reported by UMIST for \ce{CO}, originally from \citet{Black1975}, where they report $c_{\ce{CO},T} = 3$ whereas we find $c_{\ce{CO},T} = 1.943$. We report the mean coefficients in Table \ref{tab:coeffs}, where the mean is taken of the coefficients for total hydrogen nuclei column densities ranging from $10^{20}$ -- $10^{23}$ cm$^{-2}$. We note that these reaction rates are tailored for molecular cloud-like environments, e.g. those with a prescribed external cosmic-ray spectrum matching the ``High'' model from \citet{Padovani2022} (see also \citealt{Ivlev2015}), and have total hydrogen nuclei column densities between $10^{20} \leq N_{\rm H} \leq 10^{23}$ cm$^{-2}$.

There is a substantial difference in the reaction rates as found using the SCAR and BEB methods. While the additivity rules account for the {\it geometric} overlap of the atomic cross sections, it does not account for the structure of the molecule orbitals nor differences in eBEs due to molecular bonds. Therefore, we do not recommend the use of additivity rules to compute the cross sections for use in astrochemical modeling, unless it is impractical to compute molecular orbitals and use the BEB cross section.

\begin{figure}
    \centering
    \includegraphics[width=0.5\textwidth]{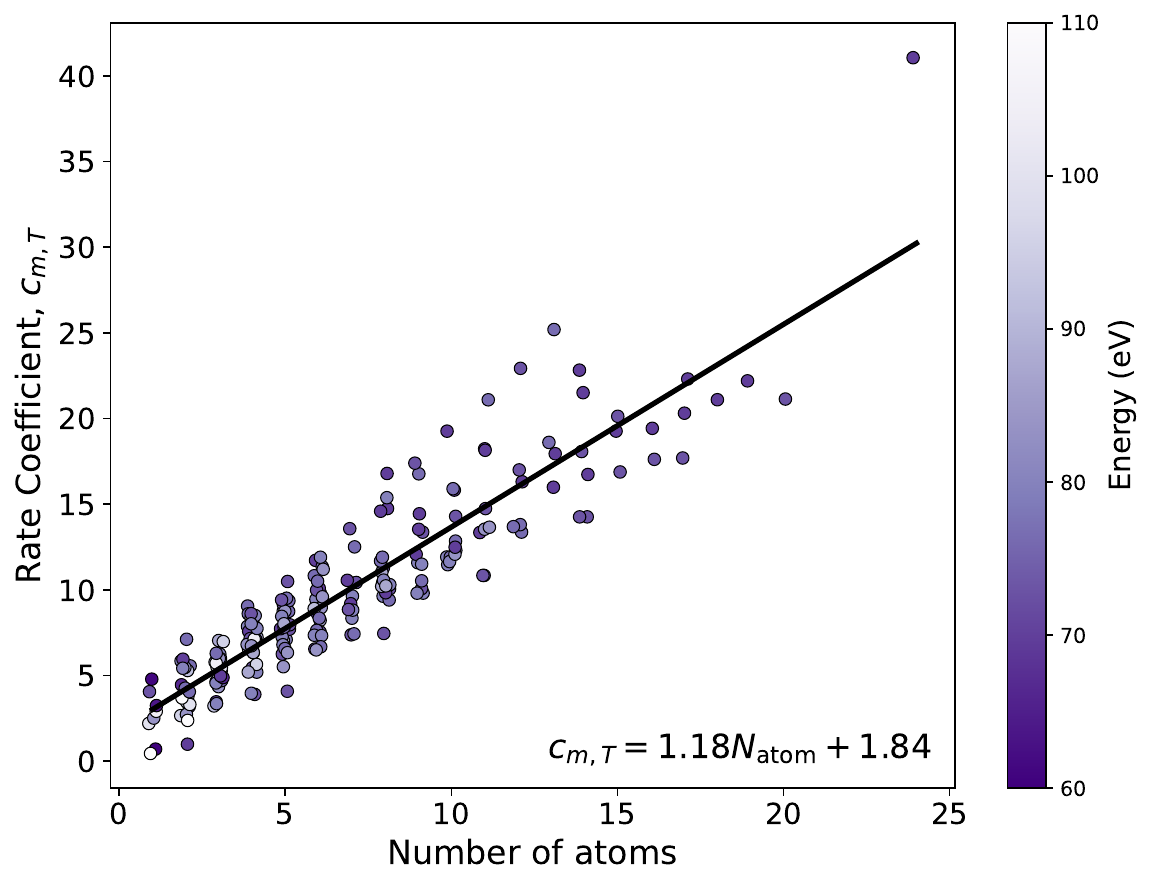}
    \caption{\label{fig:reactNatom}Total ionization rate coefficient, $c_{m,T}$, as a function of the constituent number of atoms in each molecule. The color denotes the energy at the cross section maximum.}
\end{figure}

Figure \ref{fig:reactNatom} shows the mean $c_{m, T}$ coefficient as a function of the number of atoms. Similar to Figure \ref{fig:maxXS}, we find a general increase in the reaction rate with the number of constituent atoms. The figure also shows a best-fit relation
\begin{equation}
    c_{m,T} = 1.19 {\rm N_{atom}} + 1.84,
\end{equation}
where ${\rm N_{atom}}$ is the number of constituent atoms. The relationship reproduces the general trend for ${\rm N_{atom}} \leq 20 $. We also fit the coefficients as a function of the number of valence electrons, which produces a slightly better fit
\begin{equation}
    c_{m,T} = 0.45 {\rm N_{v,elec}} - 0.64,
\end{equation}
where ${\rm N_{v,elec}}$ is the number of valence electrons in the molecule. This fit was constrained only for molecules with $9 \leq {\rm N_{v,elec}} \leq 70$. In both cases, there is significant scatter of approximately a factor of 2 around the fit trends, so caution should be used if using these for molecules not in the database. 

\longtab{
\begin{longtable}{cc}
\caption{\label{tab:coeffs}Table of the chemical reaction rates, $c_{m,T}$}\\
\hline\hline
Atom & $c_{m, T}$ \\
\hline
\endfirsthead
\caption{continued.}\\
\hline\hline
Molecule & $c_{m, T}$ \\
\hline
\endhead
\hline
\endfoot
\ce{H} & 0.713 \\
\ce{He} & 0.459 \\
\ce{C} & 3.252 \\
\ce{N} & 2.522 \\
\ce{O} & 2.208 \\
\ce{P} & 4.796 \\
\ce{S} & 4.064 \\
\ce{Ar} & 2.919 \\
\hline
Diatomic & $c_{m, T}$ \\
\hline
\ce{C2} & 4.061 \\
\ce{CF+} & 1.383 \\
\ce{CH} & 3.190 \\
\ce{CH+} & 0.795 \\
\ce{CN} & 3.881 \\
\ce{CN-} & 28.653 \\
\ce{CO} & 3.476 \\
\ce{CO+} & 1.401 \\
\ce{CP} & 5.960 \\
\ce{CS} & 5.432 \\
\ce{H2} & 1.000 \\
\ce{HF} & 2.388 \\
\ce{HS} & 4.271 \\
\ce{N2} & 3.319 \\
\ce{NH} & 2.740 \\
\ce{NO} & 3.779 \\
\ce{NO+} & 1.316 \\
\ce{NS} & 5.474 \\
\ce{O2} & 3.699 \\
\ce{OH} & 2.666 \\
\ce{OH+} & 0.809 \\
\ce{PH} & 4.465 \\
\ce{PN} & 5.570 \\
\ce{PO} & 5.855 \\
\ce{S2} & 7.123 \\
\ce{SO} & 5.300 \\
\hline
Triatomic & $c_{m, T}$ \\
\hline
\ce{C2H} & 4.987 \\
\ce{C3} & 6.306 \\
\ce{CCN} & 6.004 \\
\ce{CCO} & 5.990 \\
\ce{CH2} & 3.473 \\
\ce{CNO} & 5.790 \\
\ce{CO2} & 5.287 \\
\ce{H2O} & 3.232 \\
\ce{H2O+} & 1.028 \\
\ce{H2S} & 4.710 \\
\ce{H3+} & 0.306 \\
\ce{HCN} & 4.357 \\
\ce{HCO} & 4.662 \\
\ce{HCO+} & 1.643 \\
\ce{HNC} & 4.574 \\
\ce{HNO} & 4.783 \\
\ce{HO2} & 4.792 \\
\ce{HOC+} & 1.705 \\
\ce{HPO} & 6.258 \\
\ce{N2H+} & 1.622 \\
\ce{N2O} & 5.460 \\
\ce{NH2} & 3.366 \\
\ce{NO2} & 5.517 \\
\ce{O3} & 5.734 \\
\ce{OCN} & 5.625 \\
\ce{OCS} & 7.049 \\
\ce{PH2} & 4.887 \\
\ce{SO2} & 6.986 \\
\hline
4-atomic & $c_{m, T}$ \\
\hline
\ce{C2H2} & 5.463 \\
\ce{C2N2} & 7.237 \\
\ce{C3N} & 7.753 \\
\ce{C3N-} & 35.329 \\
\ce{C3O} & 8.028 \\
\ce{C4} & 8.618 \\
\ce{CH2O} & 5.212 \\
\ce{CH3} & 3.906 \\
\ce{CHNO} & 6.348 \\
\ce{CO2H+} & 2.713 \\
\ce{H2CN} & 5.206 \\
\ce{H2CS} & 7.162 \\
\ce{H2O2} & 5.657 \\
\ce{H3O+} & 1.281 \\
\ce{HC2O} & 6.741 \\
\ce{HCCN} & 6.777 \\
\ce{HCCO} & 6.820 \\
\ce{HCNH+} & 1.924 \\
\ce{HCNO} & 6.692 \\
\ce{HCNS} & 9.062 \\
\ce{HNCO} & 6.345 \\
\ce{HNCS} & 8.507 \\
\ce{HOCO} & 6.558 \\
\ce{HSCN} & 7.877 \\
\ce{HSSH} & 8.610 \\
\ce{NCSH} & 7.878 \\
\ce{NH3} & 3.971 \\
\ce{NO3} & 7.138 \\
\ce{SO3} & 8.382 \\
\ce{c-C3H} & 7.137 \\
\ce{l-C3H} & 7.579 \\
\ce{l-C3H+} & 2.881 \\
\hline
5-atomic & $c_{m, T}$ \\
\hline

\ce{C2H2O} & 7.272 \\
\ce{C2H3} & 6.257 \\
\ce{C2HNO} & 8.029 \\
\ce{C4H} & 9.160 \\
\ce{C4H-} & 47.898 \\
\ce{C5} & 10.496 \\
\ce{CH2NH} & 6.580 \\
\ce{CH2OH} & 6.342 \\
\ce{CH2PH} & 7.740 \\
\ce{CH2SH} & 7.959 \\
\ce{CH3O} & 5.522 \\
\ce{CH3S} & 7.711 \\
\ce{CH4} & 4.090 \\
\ce{H2CCN} & 7.107 \\
\ce{H2CCS} & 9.419 \\
\ce{H2NCO+} & 4.804 \\
\ce{HC2NC} & 8.750 \\
\ce{HC3N} & 8.461 \\
\ce{HC3O} & 8.760 \\
\ce{HCCCO} & 8.768 \\
\ce{HCNCC} & 9.519 \\
\ce{HCOOH} & 7.030 \\
\ce{HNC3} & 9.371 \\
\ce{N2H3} & 6.805 \\
\ce{NCCNH+} & 3.890 \\
\ce{NH2CN} & 7.681 \\
\ce{NH4+} & 1.469 \\
\ce{NHCNH} & 7.828 \\
\ce{c-C3H2} & 7.969 \\
\ce{l-C3H2} & 8.009 \\
\hline
6-atomic & $c_{m, T}$ \\
\hline

\ce{C2H4} & 6.552 \\
\ce{C3H3} & 8.596 \\
\ce{C4H2} & 10.103 \\
\ce{C5H} & 11.728 \\
\ce{C5N} & 11.910 \\
\ce{C5N-} & 41.216 \\
\ce{CH2CCO} & 9.963 \\
\ce{CH2NH2} & 7.559 \\
\ce{CH3CN} & 7.360 \\
\ce{CH3NC} & 7.644 \\
\ce{CH3NCO} & 10.829 \\
\ce{CH3NH} & 6.685 \\
\ce{CH3OH} & 6.502 \\
\ce{H2CCNH} & 8.355 \\
\ce{HC2CHO} & 9.454 \\
\ce{HC3NH+} & 4.418 \\
\ce{HCOCHO} & 8.916 \\
\ce{HNCHCN} & 8.976 \\
\ce{HNCHSH} & 9.831 \\
\ce{N2H4} & 7.337 \\
\ce{NC4N} & 11.206 \\
\ce{NH2CHO} & 8.228 \\
\ce{NH2CHS} & 10.516 \\
\ce{c-H2C3O} & 9.600 \\
\ce{l-HC4H} & 9.993 \\
\ce{l-HC4N} & 11.367 \\
\hline
7-atomic & $c_{m, T}$ \\
\hline
\ce{C2H5} & 7.383 \\
\ce{C6H} & 13.576 \\
\ce{C6H-} & 50.847 \\
\ce{CH2CCH2} & 9.199 \\
\ce{CH2CHCN} & 9.646 \\
\ce{CH3C2H} & 8.860 \\
\ce{CH3CHO} & 8.352 \\
\ce{CH3CHS} & 10.436 \\
\ce{CH3NH2} & 7.440 \\
\ce{H2CCHOH} & 8.824 \\
\ce{HC5N} & 12.515 \\
\ce{c-C2H4O} & 10.563 \\
\hline
8-atomic & $c_{m, T}$ \\
\hline
\ce{C2H3NH2} & 10.056 \\
\ce{C2H6} & 7.462 \\
\ce{C3H5} & 9.836 \\
\ce{C4H4} & 11.289 \\
\ce{C6H2} & 14.753 \\
\ce{C7H} & 16.800 \\
\ce{CH2CCHCN} & 11.912 \\
\ce{CH2CHCHO} & 10.575 \\
\ce{CH2OHCHO} & 10.422 \\
\ce{CH3C3N} & 11.691 \\
\ce{CH3CHNH} & 9.417 \\
\ce{CH3COCN} & 11.037 \\
\ce{CH3COOH} & 10.210 \\
\ce{CH3OCH2} & 9.636 \\
\ce{HCOOCH3} & 10.238 \\
\ce{NC6N} & 15.387 \\
\ce{NH2CH2CN} & 10.308 \\
\ce{l-HC6H} & 14.597 \\
\hline
9-atomic & $c_{m, T}$ \\
\hline
\ce{C2H5CN} & 10.534 \\
\ce{C2H5OH} & 9.813 \\
\ce{C4H5} & 12.083 \\
\ce{C5H4} & 13.373 \\
\ce{C8H} & 17.401 \\
\ce{C8H-} & 54.738 \\
\ce{CH2CCCCH2} & 14.448 \\
\ce{CH3C4H} & 13.542 \\
\ce{CH3CONH2} & 11.505 \\
\ce{CH3NHCHO} & 11.590 \\
\ce{CH3OCH3} & 9.817 \\
\ce{HC7N} & 16.778 \\
\ce{l-C3H6} & 10.079 \\
\hline
10-atomic and more & $c_{m, T}$ \\
\hline
\ce{C10H2} & 22.934 \\
\ce{C2H5CHO} & 11.476 \\
\ce{C2H5OCH3} & 13.377 \\
\ce{C3H5CN} & 12.852 \\
\ce{C3H7} & 10.849 \\
\ce{C3H8} & 10.849 \\
\ce{C4H10} & 14.267 \\
\ce{C4H7} & 13.353 \\
\ce{C4H8} & 14.267 \\
\ce{C5H5} & 14.297 \\
\ce{C5H6} & 14.750 \\
\ce{C5H7} & 16.314 \\
\ce{C5H8} & 15.996 \\
\ce{C5H9} & 16.741 \\
\ce{C6H4} & 15.829 \\
\ce{C6H7} & 17.953 \\
\ce{C6H8} & 18.079 \\
\ce{C7H4} & 18.154 \\
\ce{C7H7} & 22.830 \\
\ce{C8H2} & 19.269 \\
\ce{C8H6} & 21.514 \\
\ce{CH2CHCHCH2} & 12.495 \\
\ce{CH2OHCH2OH} & 12.313 \\
\ce{CH3C5N} & 15.909 \\
\ce{CH3C6H} & 18.239 \\
\ce{CH3COCH3} & 11.656 \\
\ce{CH3COOCH3} & 13.661 \\
\ce{CH3OCH2OH} & 12.089 \\
\ce{HC11N} & 25.198 \\
\ce{HC9N} & 21.100 \\
\ce{HCOOC2H5} & 13.549 \\
\ce{R-CH3CHCH2O} & 11.925 \\
\ce{S-CH3CHCH2O} & 11.925 \\
\ce{c-C6H5CN} & 18.613 \\
\ce{c-C6H6} & 17.007 \\
\ce{i-C3H7CN} & 13.695 \\
\ce{n-C3H7CN} & 13.813 \\
\ce{C14H10} & 41.065 \\
\ce{C5H10} & 16.887 \\
\ce{C5H11} & 17.624 \\
\ce{C5H12} & 17.702 \\
\ce{C6H10} & 19.434 \\
\ce{C6H11} & 20.318 \\
\ce{C6H12} & 21.104 \\
\ce{C6H13} & 22.204 \\
\ce{C6H14} & 21.143 \\
\ce{C6H9} & 19.272 \\
\ce{C7H10} & 22.314 \\
\ce{C7H8} & 20.136 \\
\end{longtable}
}


\subsection{Database description and file format}
We include a number of different formatted files for the total electron-impact cross sections, ionization potentials, molecular and atomic orbitals, and reaction rate coefficients. Cross sections are found in the \verb|ion_xs/| directories for the SCAR, HF, MP2 and CCSD(T) calculations and atoms. cross-section files are formatted as in the following example, {\tt co.xs}
{\small 
\begin{verbatim}
 # CO cross section
 256
 1.00000e+01     0.00000e+00     0.00000e+00
 1.04618e+01     0.00000e+00     0.00000e+00
 1.09450e+01     0.00000e+00     0.00000e+00
 1.14505e+01     0.00000e+00     0.00000e+00
 1.19793e+01     0.00000e+00     0.00000e+00
 1.25325e+01     0.00000e+00     0.00000e+00
 1.31113e+01     0.00000e+00     0.00000e+00
 1.37169e+01     0.00000e+00     0.00000e+00
                   {more data}
 9.13659e+05     8.52230e-03     6.30487e-03
 9.55855e+05     8.17497e-03     6.04801e-03
 1.00000e+06     7.84171e-03     5.80155e-03
\end{verbatim}
}
\noindent where the first row gives the number of electron energy bins. The later rows give the data in (two) three columns. For BEB cross sections, the columns provide electron impact energy (eV), BEB cross section in units of $\sigma_0^2$ and, if available, the damped-BEB cross section in units of $\sigma_0^2$. For SCAR cross sections, the three columns provide energy (eV) and SCAR cross section in units of $\sigma_0^2$.

The orbital information is stored in two directories containing the orbitals in the NIST format, \verb|NIST_orbitals/| and including all orbitals, \verb|full_orbitals/|. An example of the NIST formatted file, {\tt co.norb}
{\small 
\begin{verbatim}
 #CO NIST Orbitals
 #Orbital      B            U      N Q
 1        562.416986   794.326705  2	1
 2        309.246847   436.294142  2	1
 3         41.332626    78.262641  2	1
 4         21.888866    71.857734  2	1
 5         17.015000    53.945104  2	1
 6         17.015000    53.945104  2	1
 7         14.266000    42.788163  2	1
\end{verbatim}
}
\noindent where the B column gives the eBE (eV), U gives the average electron kinetic energy (eV), N gives the orbital occupation number and Q is a scaling factor to include higher-order ionization effects. We also show an example of the full orbital file, {\tt co.forb}
{\small 
\begin{verbatim}
 #CO Full Orbitals
 #Alpha orbitals
 #Orb.         KE       BE-HF   BE-P3+     Pole
    1   794.32671   562.41699      ***      ***
    2   436.29414   309.24685      ***      ***
    3    78.26264    41.33263      ***      ***
    4    71.85773    21.88887      ***      ***
    5    53.94510    17.35920   17.015    0.903
    6    53.94510    17.35920   17.015    0.903
    7    42.78816    15.11922   14.266    0.910
\end{verbatim}
}
\noindent where the file will give the information for both alpha and beta orbitals, for open-shell molecules that were computed with an unrestricted formalism. The KE and BE-HF columns give the average electron kinetic energy and eBE (eV) computed by the population analysis in {\sc Gaussian16} following geometry optimization. The latter 2 columns include the results from EPT, where {\tt ***} denotes orbitals where EPT was not computed. For orbitals where it was computed, we include the EPT-corrected HF canonical orbital energies (at P3+ level) results and the pole strength (PS). 

We include the computed ionization potentials in the \verb|ips/| directory for the compiled NIST and computed CAM-B3LPY and CCSD(T) molecules. These are two-column files with the molecule and ionization potential in eV. We also include two network files containing all the new molecular and atomic ionization rate coefficients in the \verb|networks/| directory. These network files are in the KIDA and UMIST formats, {\tt alecs.kida.in} and {\tt alecs.umist.d}, respectively. We recommend users to only include ionization rates for molecules in which there are associated recombination rates for the ion. In our rate files, we assume all ionizations occur in the manner \ce{AB + e- -> AB+ + e- + e-}, and emphasize potential users should check their  chemical networks for other branches such as \ce{AB + e- -> A + B+ + e- + e-}. We leave such differences for future database releases. Finally, in the \verb|geoms/| folder we include all molecule geometries computed at the HF, MP2, and CCSD(T) levels. The MP2 and CCSD(T) geometries are formatted as {\sc Pdb} files and the HF geometries as {\sc Xyz}.

\subsection{Astrochemical Modeling}

We include the new calculations in a model that uses the \textsc{KIDA} reactions framework, providing a zero-dimensional model as in Sect.\,3.1 of \citet{Wakelam2015}. It is important to note that, within our context, this test has not been designed to quantify the impact of the new rates in an astrophysical environment, but to show how the new rate equations are compatible with previously established formats. Therefore, the cross sections may play a more important role in, e.g., ice chemistry, disks, planetary atmospheres, exomoons, or cometary environments. However, this analysis is beyond the aims of the present paper, and it will be discussed in a forthcoming work.

We model a gas with total hydrogen density $n_{\rm H,tot}=2\times10^{4}$\,cm$^{-3}$, gas temperature $T=10$\,K, initial conditions as in Tab.\,\ref{tab:initial_conditions}, and no dust (e.g., see \citealt{Hincelin2011,Loison2014}).
We compute the ionization rate $\zeta(N_{\ce{H2}})$ including protons, primary and secondary electrons, and secondary from primary electrons. Note that the ionization rate is a function of the column density $N$ (see Eq.\,\ref{eq:coeffs}), and that we assume that the visual extinction (necessary for the photochemistry) is $A_V=1.0638\times10^{-21}\,N$, where $N$ is in units of cm$^{-2}$.
To model the time-dependent evolution of the chemical abundances for 10\,Myr, we make use of \textsc{kmarx} (Grassi in prep., commit \verb+58f6ac9+), a Python-based database that allows us to solve the chemical ordinary differential equations with a standard BDF solver\footnote{\url{https://github.com/Nicholaswogan/NumbaLSODA}} \citep{Hindmarsh1992}. We include the HTML output produced by the code in the database as a zip file in the \verb|chem_models/| folder, where the reactions present in this work are listed under the class \texttt{CRReactionAdv}. 

When using our chemical network, we replace the ionization reactions from KIDA where possible (i.e., H, He, N, O, H$_2$, and CO) and add the missing ionization rates that will impact the molecules listed in Tab.\ref{tab:initial_conditions}. We do not add the species that creates sinks or sources in the chemical network, i.e., species that only appear in the products or in the reactants, respectively.
We also note that, for the sake of comparison, the cosmic-ray reaction rates from KIDA are scaled so that the H$_2$ ionization matches the ionization rate of molecular hydrogen in our database. The aim of this scaling is to avoid discrepancies determined by different assumptions in the cosmic-ray spectra employed.

\begin{figure}
    \centering
    \includegraphics[width=0.5\textwidth]{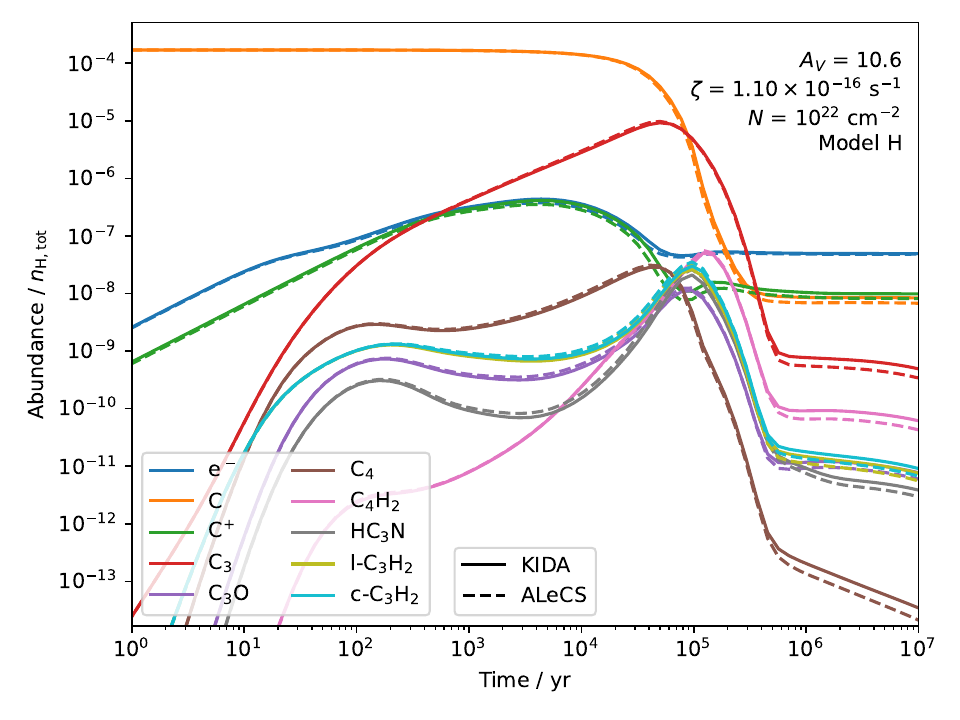}
    \caption{\label{fig:evolution}Comparison of the chemical evolution using KIDA reaction rates (solid lines) and ours (dashed lines). We report the species that present a difference of at least half an order of magnitude and reach at least $n(t) = 10^{-8}n_{\rm H,tot}$. The ionization rate of H$_2$ is 1.64$\times10^{-16}$\,s$^{-1}$, that for model ``High'' and with the assumptions described in the text, corresponds to $A_V$=10.6 and $N$=10$^{22}$\,cm$^{-2}$.}
\end{figure}

Fig.\,\ref{fig:evolution} shows the results obtained with the \textsc{KIDA} database reaction rates (solid lines) and calculated with the rates present in this work (dashed lines). For the sake of clarity, we only plot the species that show a difference larger than a tenth of an order of magnitude and reach at least $n(t) = 10^{-8}n_{\rm H,tot}$ during their evolution. Only a few species present a negligible discrepancy between the two databases. The extent of these variations is minimal because the critical cosmic-ray-driven reactions are very similar in ALeCS and in \textsc{KIDA}, but the results might be less interchangeable in different environments. Also, note that we have assumed that the constituents of the spectra are matching, i.e., both include protons, electrons, and secondary processes, hence the scaling mentioned above.

\begin{table}
    \centering
    \caption{\label{tab:initial_conditions}Initial abundances as a function of the total H nuclei abundance, i.e. $n_{\rm H, tot}$ \citep{Hincelin2011}. We assume $a(b)=a\times10^b$.}
    \begin{tabular}{llll}
    \hline
    Species & Abundance & Species & Abundance \\
    \hline 
    H$_2$ & 5(-1)   & P  & 2(-10) \\
    He    & 9(-2)   & N  & 6.2(-5) \\
    C     & 1.7(-4) & O  & 2.4(-4) \\
    S     & 8(-8)   & Si & 8(-9) \\
    Fe    & 3(-9)   & Na & 2(-9) \\
    Mg    & 7(-9)   & Cl & 1(-9) \\
    \hline
    \end{tabular}
\end{table}

\section{Conclusions}
We present the initial data release for the Astrochemistry Low-energy electron cross-section (ALeCS) database. In this release, we include the total ionization cross sections and ionization rate coefficients for over 200 neutral molecules of astrochemical interest calculated using three different semi-empirical methods: the SCAR \citep{Blanco2003, Blanco2010}, the BEB model \citep{Kim1994, Hwang1996} and a new dampened-BEB model (dBEB) presented here. The latter of these dampens orbitals deep within the potential well and was demonstrated to help prevent the BEB model from overestimating the ionization cross section when compared with experimental results. We also present the ionization rate coefficients, and molecular ionization rates scaled to a reference total \ce{H2} ionization rate.

The database is fully public, and will include the ionization data described above, along with the molecule orbitals and optimized geometries. In this current release, we only include the semi-empirical ionization cross sections and reference experimental data. Future releases will include more sophisticated ionization cross-section calculations along with excitation and momentum transfer. Finally, the database will include open-source software tools necessary to couple these processes to astrochemical codes. 

\begin{acknowledgements}
We thank the referee, Jonathan Tennyson, for their useful comments which improved this work. The MP2 and DFT calculations presented here were performed on the {\sc Vera} computing cluster, managed by the Chalmers Center for Computational Science and Engineering. BALG and PG are supported by the Chalmers Initiative on Cosmic Origins as Cosmic Origins Postdoctoral Fellows. TG acknowledges support of the DFG (German Research Foundation) Excellence Cluster ORIGINS - EXC-2094 - 390783311. S.B. is financially supported by ANID Fondecyt Regular (project \#1220033), and the ANID BASAL projects ACE210002 and FB210003.  CV and DH would like to thank the UK Engineering and Physical Sciences Research Council for funding via Programme Grants EP/V026690/1 and EP/T021675/1. SVG is ﬁnancially supported by VRID project 2022000507INV. GMB gratefully acknowledges support from ANID Beca de Doctorado Nacional 21200180. This is University of Texas Center for Planetary Systems Habitability (CPSH) contribution number \#0073. 
\end{acknowledgements}

\bibliographystyle{aa}
\bibliography{lib} 

\end{document}